\documentclass[reprint, superscriptaddress]{revtex4-2}
\usepackage{amsmath,amssymb}
\usepackage{stmaryrd}
\usepackage{latexsym}
\usepackage{CJK}
\usepackage{graphicx}
\usepackage{indentfirst}
\usepackage{listings}
\usepackage{xcolor}
\usepackage{booktabs}
\usepackage{stmaryrd}
\usepackage{multirow}
\usepackage{verbatim}
\usepackage{makecell}
\usepackage{lineno,hyperref}
\usepackage{longtable}
\usepackage{hyperref}
\usepackage{upgreek}
\usepackage{mathtools}
\usepackage{comment}
\usepackage{enumitem}
\usepackage[figuresright]{rotating}
\hypersetup{colorlinks,breaklinks,
            linkcolor=blue,urlcolor=blue,anchorcolor=blue,citecolor=blue}

\newcolumntype{C}[1]{>{\centering\let\newline\\\arraybackslash\hspace{0pt}}m{#1}}

\newcommand{\rmd}{\mathrm{d}}
\newcommand{\rmB}{\mathrm{B}}
\newcommand{\rmf}{\mathrm{f}}
\newcommand{\rmb}{\mathrm{b}}

\begin{document}
\title{Interface Faceting-Defaceting Mediated by Disconnections}

\author{Caihao Qiu}
\affiliation{Department of Materials Science and Engineering, City University of Hong Kong,  Hong Kong SAR, China}
\author{Marco Salvalaglio}
\affiliation{Institute of Scientific Computing, TU Dresden, 01062 Dresden, Germany}
\affiliation{Dresden Center for Computational Materials Science, TU Dresden, 01062 Dresden, Germany}
\author{David J. Srolovitz}
\affiliation{Department of Mechanical Engineering, The University of Hong Kong, Pokfulam Road, Hong Kong SAR, China}
\affiliation{International Digital Economy Academy (IDEA), Shenzhen, China}
\author{Jian Han}
\email{jianhan@cityu.edu.hk}
\affiliation{Department of Materials Science and Engineering, City University of Hong Kong, Hong Kong SAR, China}

\date{\today}

\begin{abstract}
An intrinsic feature of nearly all internal interfaces in crystalline systems (homo- and hetero-phase) is the presence of disconnections (topological line defects constrained to the interface that have both step and dislocation character). 
Disconnections play a major role in determining interface thermodynamics and kinetics.
We demonstrate that elastic interactions between disconnections lead to a thermodynamic, first-order, finite-temperature, faceting-defaceting transition, in agreement with experiments.
These elastic interactions strongly modify equilibrium interface morphologies (compared with those solely determined by anisotropic surface energy) as well as the kinetics and morphologies of migrating interfaces. 
We demonstrate these phenomena through numerical simulations based upon a general, continuum disconnection-based model for interface thermodynamics and kinetics applied to embedded particles/grains, steady-state interface migration geometries, and nominally flat interfaces.
\end{abstract}

\maketitle

\section{Introduction}\label{intro}

Most materials of technological interest are polycrystalline, have multiple phases, or both; in other words, interfaces between crystalline phases are ubiquitous. 
The interfaces separating differently oriented domains of the same crystal phase (grains) are referred to as grain boundaries (GBs), while those separating distinct crystal phases regardless of the relative crystal orientation are identified as heterophase interfaces. These interfaces may either be smoothly curved or faceted, i.e. exhibiting flat regions of fixed inclination  connected by sharp corners/edges or curved junctions. 
Varying the temperature (and other thermodynamic conditions) or inducing interface migration can lead to the faceting of smoothly curved interfaces and vice versa. 
Such faceting transitions are normally discussed in terms of the anisotropy of the interface energy (per area), $\gamma(\hat{\textbf{n}}(s))$, where $\hat{\textbf{n}}(s)$ is the interface unit normal that varies along the interface, $s$. For a closed surface, such as a particle embedded in a matrix, the interface morphology is well described by the Wulff construction \cite{1901_Wulff_theory}, i.e. the convex hull of $\gamma(\hat{\textbf{n}}(s))$, leading to  Wulff shapes. In close-to-equilibrium conditions, interfaces exhibit preferential orientations corresponding to minima of $\gamma(\hat{\textbf{n}}(s))$. They feature sharp corners for relatively large interface-energy anisotropies, and may exhibit straight facets when there are discontinuities in the derivative of the surface energy $\gamma(\hat{\textbf{n}})$ with respect to $\hat{\textbf{n}}$. 
For a nominally flat interface, faceting results in a ``hill-and-valley" or sawtooth interface profile; again, usually described in terms of discontinuities in $\gamma(\hat{\textbf{n}})$. Although the Wulff shape well describes faceted morphologies at equilibrium, out-of-equilibrium conditions can be treated analogously by focusing on the interface-normal velocity $\mathbf{v}(\hat{\textbf{n}}(s))$ and similar constructions (leading to kinetic Wulff shapes)
\cite{1958_frank,1979_shaw_jcg}.

Thermodynamic descriptions of faceting are traditionally based solely upon the minimization of the interface free energy and the Wulff construction ~\cite{1951_Herring_surface_theo,1964_Cabrera_surface_theo}. 
Cahn described interface faceting thermodynamics and first-order faceting/defaceting transitions in terms of the inverse interface energy ($\gamma^{-1}$) plot ~\cite{1982_Cahn_gb_theo}. 
At low temperatures, the interface energy vs. interface inclination exhibits minima as ``cusps'', usually associated with atomically closed-packed, low-energy interfaces. With increasing temperature, however, these cusps become shallow, smooth, and/or disappear, leading to defaceting and smoothly curved interface morphologies ~\cite{1984_Rottmandefacet_exp}.
Although interface energy anisotropy can explain many commonly observed interface faceting-defaceting phenomena~\cite{2020_Priedeman_capillarity_exp, 2012_Kang_facetcapillarity_MD, 2008_Kirch_facetcapillarity_halfloop_exp}, experimental observations demonstrate that shear stress can also drive the faceting transformations~\cite{2005_Hanyu_stressfacet_exp}.

As is now widely accepted, interface migration is mediated by disconnections; these are line defects possessing step and/or dislocation character~\cite{2020_Hirth_disconnection_review, 2018_Han_gbkinetics_review, 2013_Rajabzadeh_disconnection_exp}. 
The dislocation character of disconnections implies that interfaces have elastic fields which  influence both interface migration and interface faceting-defaceting (as  demonstrated below).
The question addressed here is how, and to what extent, these elastic interactions modify faceting-defaceting behavior and interface mobility relative to the classical (anisotropic), capillarity-based theory (i.e., pure capillarity).
Some researchers~\cite{2003_Hamilton_stresseffect_theo, 2009_Wu_stresseffect_MD} analyzed the scale of faceting by associating dislocation character with facet junctions based on atomistic and continuum elastic approaches. 
They concluded that the stresses produced by the intrinsic junction dislocation character were too small to counter interfacial tension and maintain finite length facets ~\cite{2003_Hamilton_stresseffect_theo, 2009_Wu_stresseffect_MD}. 
Recently, Medlin et al.~\cite{2017_Medlin_stresseffect_MD} extended this analysis to include elastic interactions between intrinsic junctions and secondary GB dislocations/disconnections (i.e., GB disconnections of large Burgers vector). They concluded that GB faceting is strongly influenced by secondary GB dislocations/disconnections, even for GBs with small inclination angles. 
This study did not, however, examine the wide range of phenomena associated with faceting transitions, the effect of external stress, interface migration, or their influence on microstructure evolution. 
Here, we propose a thermodynamic and kinetic model of interface faceting based upon the crystallographic structure of crystalline interfaces that includes both capillarity (interface energy) and elasticity effects. 

We recently proposed a continuum equation of motion for arbitrarily curved interfaces based upon disconnection glide~\cite{2022_Han_EOM_theory, 2022_Marco_diffuse_PF}. 
In this study, we investigate faceting-defaceting behavior, interface mobility, and interface morphology evolution through a series of numerical and theoretical approaches based upon disconnection migration. 
We compare these results with  experimental observations and conclude that disconnection motion is key to unifying a wide range of phenomena associated with faceting-defaceting and interface morphology evolution. 
These results  apply to grain boundaries and heterophase interfaces in most crystalline systems.

\section{Interface Bicrystallography and Equation of Motion}

\begin{figure}[t]
\includegraphics[width=0.95\linewidth]{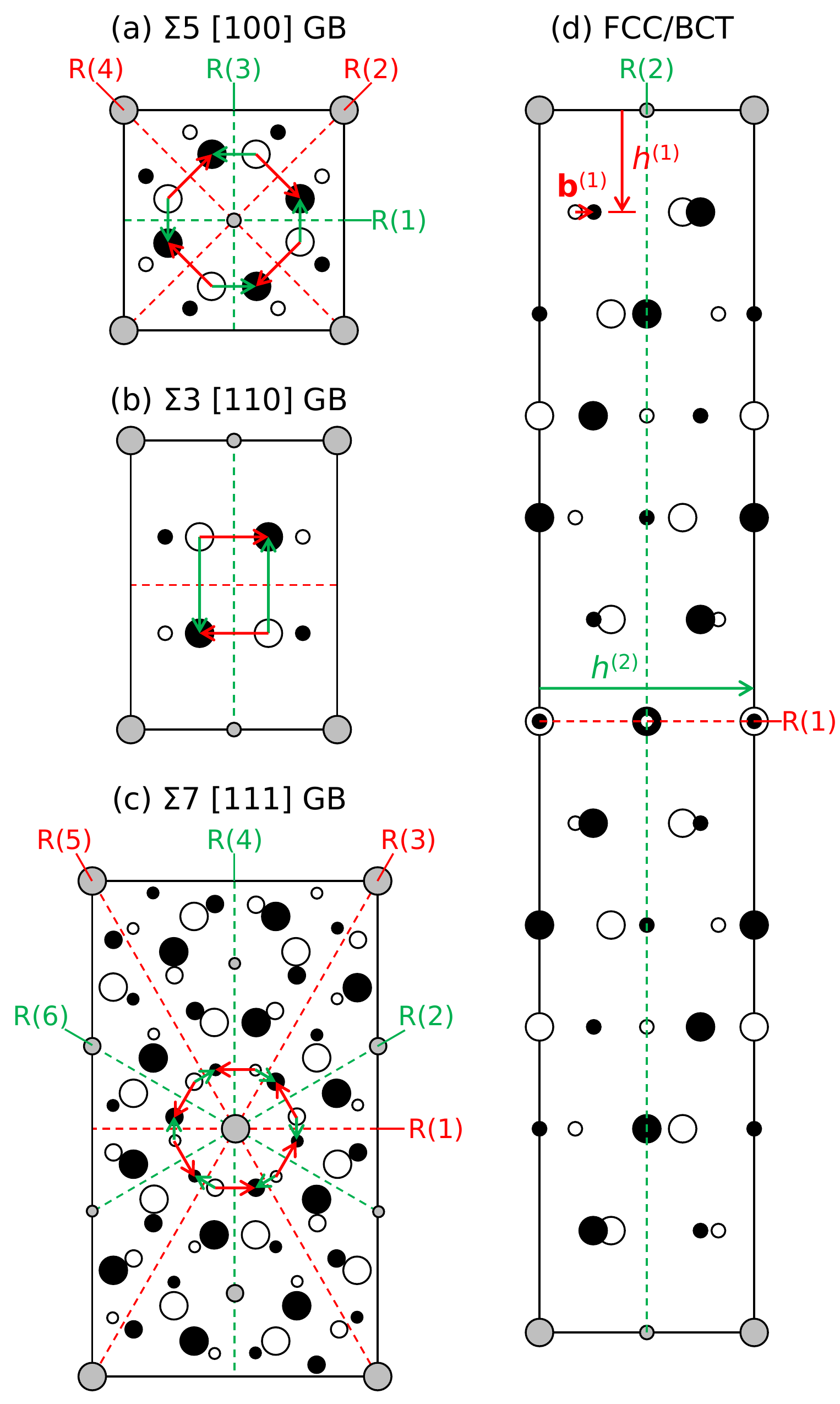}\hspace{-1.78em}%
\caption{Dichromatic patterns for two crystals (white and black) with different orientations or lattice structures. 
(a) $\Sigma5$ $[100]$, (b) $\Sigma3$ $[110]$ and (c) $\Sigma7$ $[111]$ misoriented FCC crystals. 
(d) Superimposed FCC (white) and BCT (black) crystals with the orientation relationship: $(111)_\text{FCC} \parallel (101)_\text{BCT}$ $[2\bar{1}\bar{1}]_\text{FCC} \parallel [10\bar{1}]_\text{BCT}$. 
The gray points form the CSL and are  coincident sites of the white and black crystals. 
The dash lines denote  reference interface planes. 
In (d), the arrows labeled  $\mathbf{b}^{(1)}$ and $h^{(1)}$ are the Burgers vector and step height of the disconnection on the reference interface indicated by the red dashed line. 
The arrow $h^{(2)}$ is the step height of the disconnection (pure step) on the reference interface indicated by the green dashed line.
}
\label{fig_bicrystallography}
\end{figure}

Consider two crystal lattices (differently oriented and/or of different crystal structures) meeting at an interface.
Extend the two lattices to fill the whole space such that they interpenetrate to form a dichromatic pattern. 
Lattice points of two lattices coincide at some sites in the dichromatic patters to form the coincidence-site lattice (CSL)~\cite{1966_Ranganathan_CSL_theory, 1966_Bollmann_CSL_theory, 1997_Hirth_CSL_theory}. 
Figure~\ref{fig_bicrystallography} shows the dichromatic patterns associated with the formation of several GBs: a $\Sigma5$ $[100]$ GB, a $\Sigma3$ $[110]$ GB, and a $\Sigma7$ $[111]$ GB in face-centered cubic (FCC) crystals. 
Figure~\ref{fig_bicrystallography}d shows the pattern formed by interpenetrating FCC and body-centered tetragonal (BCT) crystals, which can be used to describe a heterophase interface, e.g., the interface between an FCC parent phase and a BCT martensite in Ni$_2$MnGa.   
In Fig.~\ref{fig_bicrystallography},  sites in one lattice are shaded  black and in the other white; the coincidence sites are  gray~\cite{2018_Han_gbkinetics_review}.

To form an interface, select a plane in the CSL, remove one lattice on one side of the plane and remove the other lattice on the other side. 
We choose  close-packed planes of the CSL as  reference interface planes, indicated by the dashed lines in Fig.~\ref{fig_bicrystallography}.
Reference interfaces are highly coherent and usually correspond to  cusps in the interface energy as a function of macroscopic degrees of freedom (facet planes usually correspond to such reference planes). 
Multiple reference planes exist for each dichromatic pattern. 
For example, the $\Sigma5$ $[100]$ pattern possesses four references, i.e., the dashed lines in Fig.~\ref{fig_bicrystallography}a. 
For an $n$-reference pattern, ``R($k$) interface'' refers to the reference interface with inclination $\phi^{(k)} = k\pi/n$ ($k \in [0, n-1]$).

The line defect on each reference interface (disconnection) has an associated  step height and/or Burgers vector. 
The displacement-shift-complete (DSC) lattice consist of the set of all displacements of one lattice relative to the other that conserve the dichromatic pattern. 
Disconnection Burgers vectors correspond to translation vectors of the DSC lattice. 
For example, for the FCC/BCT pattern shown in Fig.~\ref{fig_bicrystallography}d, the red and green dashed lines, respectively, as R(1) and R(2) interface planes; the Burgers vector and step height for  disconnections on R(1)  are $\mathbf{b}^{(1)}$ and $h^{(1)}$. 
For the disconnection on the R(2) interface the the Burgers vector is zero and the step height is $h^{(2)}$ (i.e.,  this disconnection is a  pure step). 

An interface, slightly inclined with respect to a reference plane, can be represented as the reference interface with a superimposed distribution of disconnections with step character. 
Hence, an arbitrarily curved interface may be described as a distribution of disconnections on the  reference interfaces. 
The disconnections on each segment of an interface are those of the two reference interfaces most closely aligned with the local, inclined  segment; see Ref.~\cite{2022_Han_EOM_theory} and Supplementary Material (SM).

The conventional driving force for interface migration is capillarity (i.e., curvature-driven interface migration), which may be described in terms of the motion of pure-step disconnections  (no Burgers vector). 
However, the disconnections associated with general interfaces (e.g., those observed during microstructure evolution and faceting-defaceting)  usually have both step and dislocation character. 
Elastic interaction between disconnection Burgers vectors  also play an important role in microstructure evolutions and interface faceting-defaceting.
In our continuum model, the equation of motion for an arbitrarily curved interface may be expressed as \cite{2017_Zhang_EOM_theory, 2022_Han_EOM_theory}:
\begin{equation}\label{EOM}
\mathbf{v} = \mathbf{M}
\left(\Gamma\kappa + \boldsymbol{\tau} \cdot \boldsymbol{\beta} + \psi\right) 
\hat{\mathbf{n}}, 
\end{equation}
where $\mathbf{v}$ is the velocity of an interface segment and $\mathbf{M}$ is the intrinsic mobility tensor. 
Eq.~\eqref{EOM} features three  driving forces. 
Weighted mean curvature $\Gamma \kappa$ is the capillarity driving force, with interface stiffness $\Gamma = \gamma + \gamma_{,\phi\phi}$ ($\phi$ is the local inclination angle) and local mean curvature $\kappa$.  
$\boldsymbol{\tau} \cdot \boldsymbol{\beta}$ represents the elastic driving force acting on the disconnection Burgers vectors, where $\boldsymbol{\tau}$ and $\boldsymbol{\beta}$ are the resolved shear stress and shear-coupling factor vectors of the reference interfaces. 
$\psi$ is the chemical-potential jump across the interface related to differences in the bulk free energy densities of two phases/grains meeting at the interface, accounting, for example, the driving force for grain nucleation and martensite transformations.
 
\begin{figure}[t]
\includegraphics[width=0.98\linewidth]{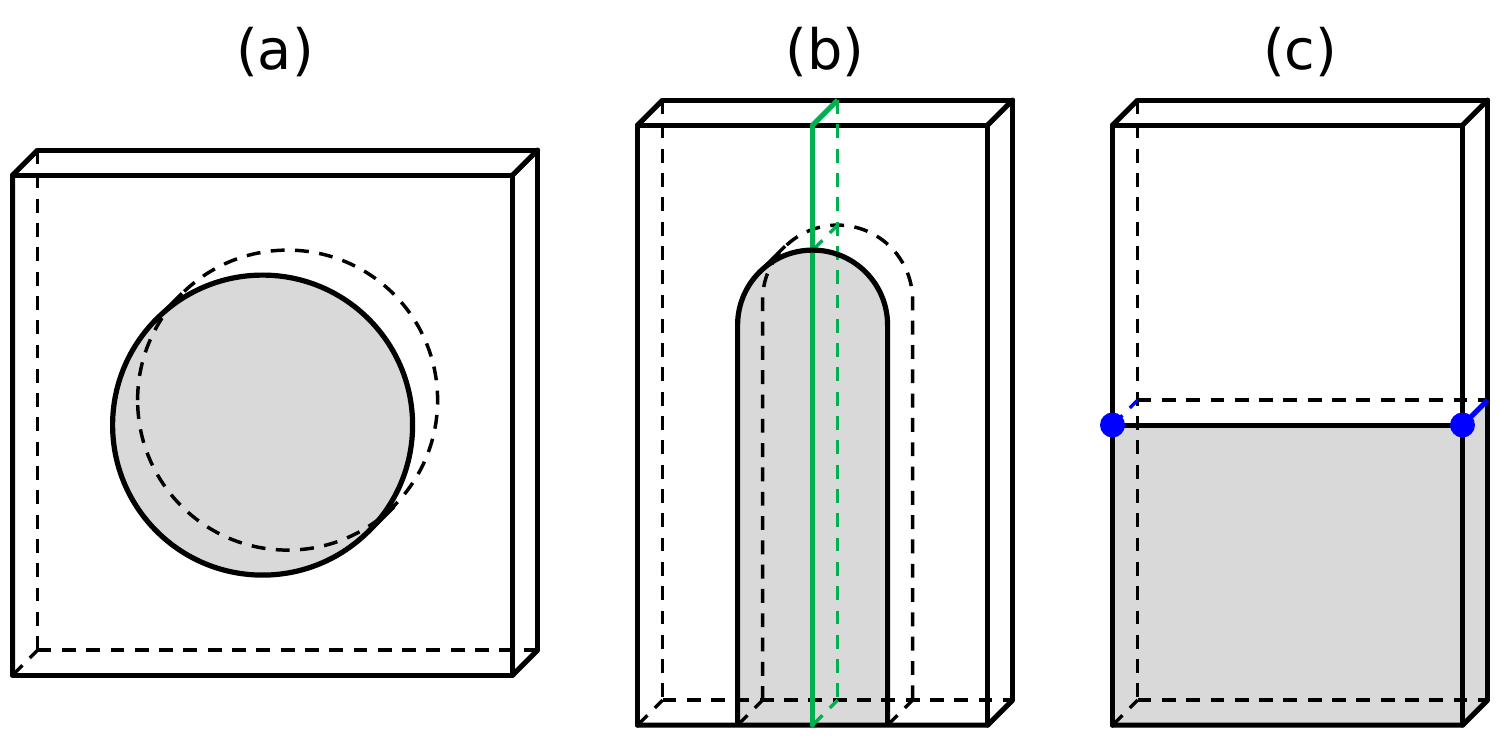}\hspace{-1.78em}%
\caption{Thin film bicrystal configurations. 
(a) Embedded grain. 
(b) GB half-loop. 
A GB quarter-loop is half of this configuration separated by the green line. 
(c) Nominally flat interface pinned at two terminations (blue lines). }
\label{configurations}
\end{figure}

In the following, we simulate the faceting-defaceting behavior for three bicrystalline thin-film configurations shown in Fig.~\ref{configurations}.
We focus on how a combination of the capillarity and elasticity associated with disconnections ($\Gamma\kappa + \boldsymbol{\tau} \cdot \boldsymbol{\beta}$) modifies the faceting-defaceting behavior based  on conventional capillarity ($\Gamma \kappa$) and/or a jump in the chemical potential at the interfaces ($\psi$).

\section{Embedded Grains}

A configuration commonly used for the study of faceting-defaceting behavior is a cylindrical grain embedded in an infinitely large matrix, as schematically shown in Fig.~\ref{configurations}a.  
Faceting-defaceting of the interface of an embedded grain is widely observed in mazed bicrystals in textured thin films. In particular, grains with continuously curved boundaries found in as-deposited thin films often facet during annealing \cite{2011_Radetic_squareAu_exp, 1996_Balluffi_squaredisconnection_exp, 2012_Radetic_squareAu_exp}. 
In this section, we focus on the effects of capillarity and elasticity on  interface faceting-defaceting of embedded grains/particles.

\begin{figure}[t]
\includegraphics[width=0.94\linewidth]{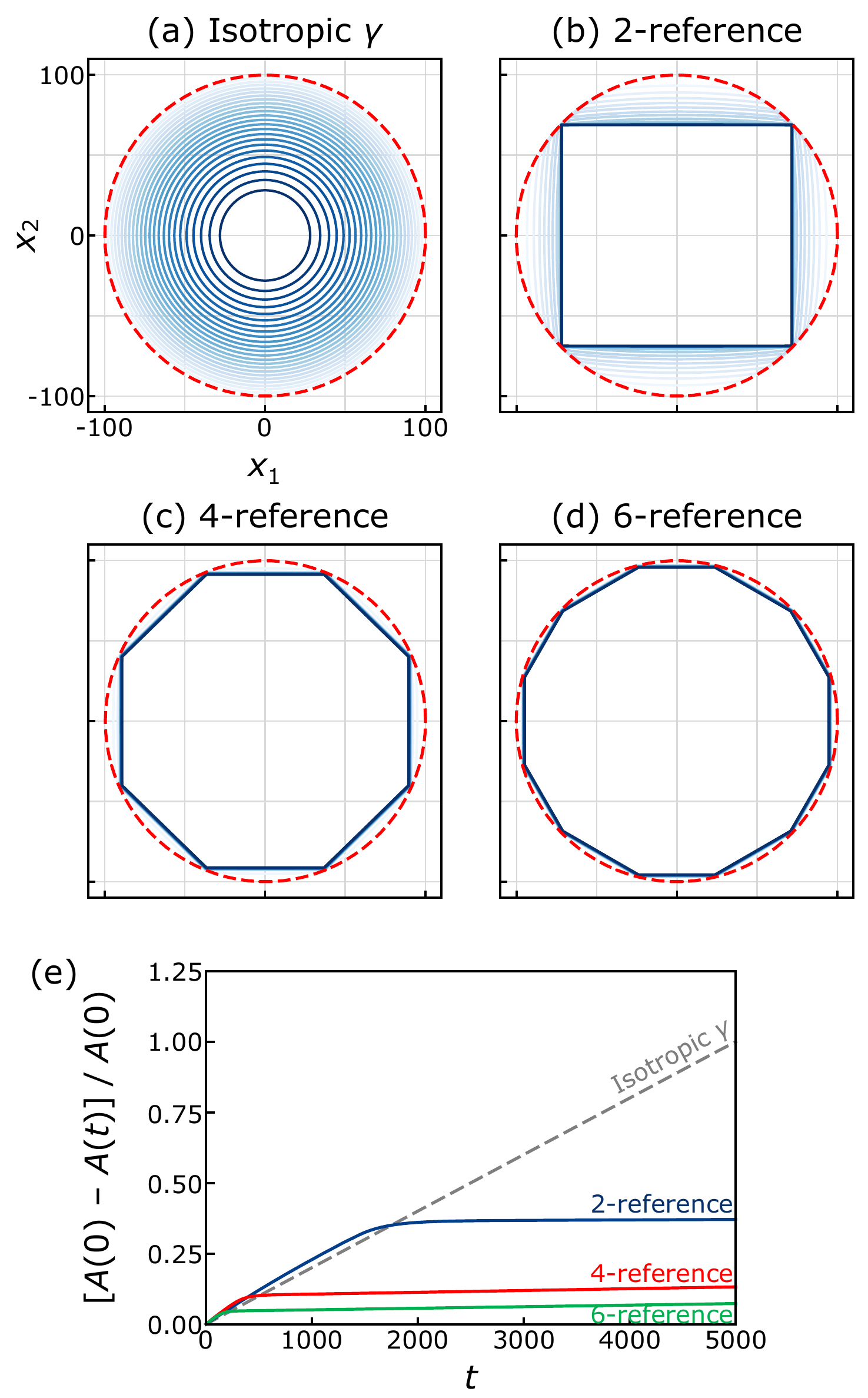}\hspace{-1.78em}%
\caption{Evolution of an initially circular domain driven by pure capillarity based on different numbers of reference interfaces according to the dichromatic pattern symmetry. (a) Isotropic interface energy. (b)-(d) 2 , 4, and 6 references. 
The red dash circle is the initial profile. 
(e) Temporal evolution of the reduced area $[A(0)-A(t)]/A(0)$.  }
\label{fig_circular_equ}
\end{figure}

We first investigate the evolving morphology of an initially circular domain in a system driven only by capillarity (weighted mean curvature) in systems with different bicrystallography (i.e., different numbers of reference interfaces) assuming isotropic interface mobility, generalizing the corresponding results of \cite{2022_Han_EOM_theory,2022_Marco_diffuse_PF}.
Figure~\ref{fig_circular_equ}a shows the evolution for the isotropic interface energy case ($\gamma = 1$). 
The red dash line is the initial grain shape, while the solid blue lines represent  interface morphologies at different stages (uniform time intervals). 
As expected, this grain shrinks as a circle and its area $A(t)$ evolves as $A(t) = A(0) - 2\pi M \gamma t$; see the gray line in Fig.~\ref{fig_circular_equ}e. 
This is classical capillarity-driven grain shrinkage.

Figure~\ref{fig_circular_equ}b-d show the evolution of  initially circular interfaces with 2, 4, and 6 reference interfaces of isotropic mobility. 
The interface energy is anisotropic~\cite{1967_Gruber_TLK_theory, 2007_McDowell_GBenergy_MD}:
\begin{equation}\label{interface_energy_phi}
\small
\gamma (\phi) = \gamma^{(1)} \left| \frac{\sin(\phi-\phi^{(1)})}{\sin(\phi^{(2)}-\phi^{(1)})}\right| + \gamma^{(2)} \left| \frac{\sin(\phi^{(2)}-\phi)}{\sin(\phi^{(2)}-\phi^{(1)})}\right|, 
\end{equation}
where the superscripts ``(1)'' and ``(2)'' denote two reference interfaces with  inclinations bounding the local interface inclination, i.e., $\phi^{(1)} \leq \phi \leq \phi^{(2)}$; $\gamma^{(1)}$ and $\gamma^{(2)}$ are the interface energies for the two reference interfaces. 
In principle, there are $n$ reference interface angles and energies, but for simplicity, we only consider the case of a single reference interface energy, $\gamma^{(1)} = \gamma^{(2)} = 1$.
This interface energy vs. inclination exhibits sharp cusps  at $\phi = k\pi /n$ ($k$ is integer).
For this anisotropic interface energy, the circle shrinks into a $2n$-sided polyhedron and then stops evolving.
The symmetry of the equilibrium shapes is consistent with the symmetry of the dichromatic pattern shown in Fig.~\ref{fig_bicrystallography} and the shape generated by the Wulff construction (see SM). 
These symmetries reflect those observed in faceted island grains for $[110]$ and $[100]$ tilt axes in FCC crystals  (e.g., see \cite{2011_Radetic_squareAu_exp, 1996_Balluffi_squaredisconnection_exp, 2012_Radetic_squareAu_exp}). 

Shrinkage stagnation of an island grain, once faceted, is due to the fact that the interface (free) energy $\gamma(\phi)$ has deep cusps at $T=0$.
At finite temperature, the configurational entropy associated with disconnection distribution must be included. 
We incorporate the temperature dependence of the interface (free) energy based upon a solid-on-solid approximation~\cite{1951_Burton_surfaceenergy_theory, 1982_Avron_SOS_theory}. 
The intrinsic interface mobility tensor $\mathbf{M}$ and the shear-coupling-factor vector $\boldsymbol{\beta}$ in Eq.~\eqref{EOM} are also temperature-dependent. 
For simplicity, we assume that all disconnection mobilities (prefactors and activation energies) are the same, such that  interface mobilities are simply Arrhenius functions with a single activation energy. 
The temperature dependences of $\mathbf{M}$ and $\boldsymbol{\beta}$ are described in  Methods  (and  SM).

\begin{figure*}[t]
\includegraphics[width=1\linewidth]{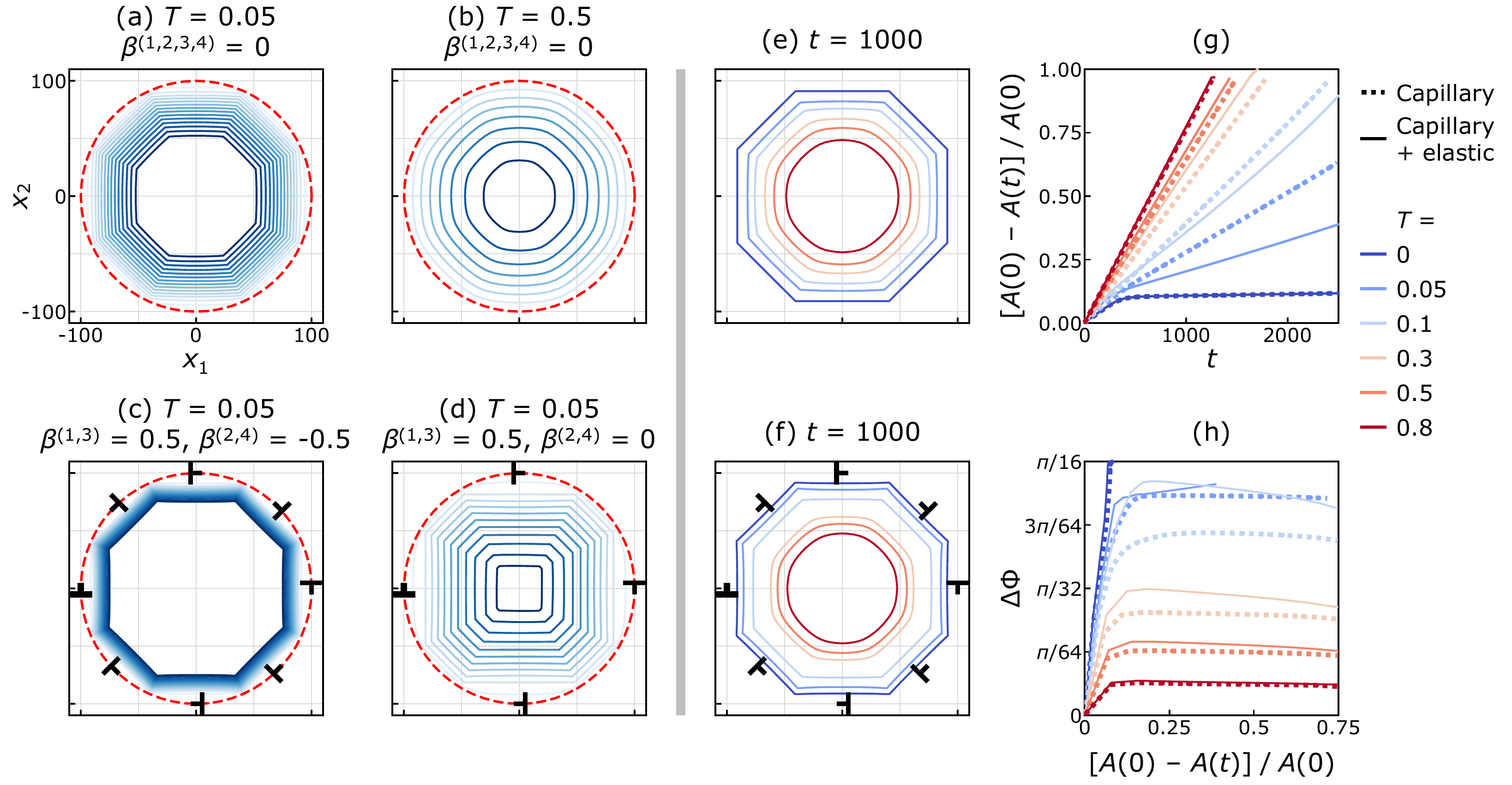}\hspace{-1.78em}%
\caption{Evolution of an initially circular interface with and without inclusion of elasticity at finite temperatures. 
(a), (b) Evolution of an initially circular interface driven by pure capillarity at temperatures of 0.05 and 0.5. 
Since the disconnections are pure steps, the shear-coupling factors for all reference interfaces are zero. 
(c), (d) Evolution of an initially circular interface driven by capillarity and elasticity at the temperature of 0.05 with different disconnection Burgers vector arrangements. 
In (c), the shear-coupling factors are $\beta^{(1)} = \beta^{(3)} = 0.5$ and $\beta^{(2)} = \beta^{(4)} = -0.5$; in (d), $\beta^{(1)} = \beta^{(3)} = 0.5$ and $\beta^{(2)} = \beta^{(4)} = 0$. 
In (a)-(d), the red dash lines are the initially circular interfaces, and the time interval is $\Delta t = 200$. 
(e), (f) Interface morphology images at different temperatures at $t = 1000$, driven by (e) pure capillarity and (f) a combination of capillarity and elasticity. 
(g) The normalized domain area $[A(t)-A(0)]/A(0)$ vs. time $t$ for 6 different temperatures. 
The dash and solid curves indicate evolution due to pure capillarity and combined capillarity and elasticity, respectively. 
(h) Degree of faceting $\Delta \Phi$, Eq.~\eqref{DPhi}, vs. normalized domain area. }
\label{fig_circular_compare}
\end{figure*}

The temperature effects on anisotropic ($n=4$) domain evolution are shown in Fig.~\ref{fig_circular_compare}.
At small, finite temperature, ( Fig.~\ref{fig_circular_compare}a), 
faceting is observed (like at $T=0$), but the embedded domain continues to shrink (unlike at $T=0$); \textit{cf.} the $T=0$ and $T=0.05$ grain evolution rates in Fig.~\ref{fig_circular_compare}g.  
At higher temperatures (Fig.~\ref{fig_circular_compare}b), the initially circular domain remains rounded (no obvious faceting) while it shrinks.
Figure~\ref{fig_circular_compare}e shows  interface morphology images at $t=1000$ for several different temperatures.
At $T=0$ the domain is fully faceted while at high temperature it is nearly circular. 
To quantify the deviation of interface morphology from the isotropic (circle) shape, we define the degree of faceting as 
\begin{equation}\label{DPhi}
\Delta \Phi 
= C- 
{\oint \min_k \left|\phi(s) -\phi^{(k)}(s)\right|\rmd s} 
\bigg/ {\oint \rmd s}, 
\end{equation}
where $s$ is the parametric coordinate along the interface curve and $C$ is a constant. 
For a closed isotropic interface, we set $C = \pi/(4n)$ such that $\Delta\Phi=0$ in the absence of faceting. 
If the interface is fully-faceted, $\Delta\Phi$ reaches the maximum $\pi/(4n)$. 
Figure~\ref{fig_circular_compare}h  shows that the degree of faceting  in the evolving shape decreases with increasing temperature.
This is indicative of a faceting-defaceting transition with increasing temperature.

Capillarity-driven evolution and the evolution controlled by the motion of disconnections with pure step character are identical~\cite{2022_Han_EOM_theory}. 
On the other hand, when disconnections have both step and dislocation character, the evolution may be very different.  
We examine the effects of such disconnections on interface faceting-defaceting; see Figs.~\ref{fig_circular_compare}c-f.
The direction of the Burgers vectors (dictated by the dichromatic pattern) in these calculations are shown as  arrows in Fig.~\ref{fig_bicrystallography}a; all  disconnections possess the same magnitude of shear-coupling factor, $\beta$.
Comparing Fig.~\ref{fig_circular_compare}a (no dislocation character) and \ref{fig_circular_compare}c (with dislocation character) shows  that including   dislocation character/elasticity reduces the interface migration velocity at $T = 0.05$
Accounting for the dislocation character  leads to sharper interface corners/vertices and makes sharper facets. 
Figure~\ref{fig_circular_compare}f shows interface morphology images for  capillarity and elasticity at $t = 1000$ for several temperatures. 
Similar to Fig.~\ref{fig_circular_compare}e,  interface corners become more rounded with increasing temperature, while the morphology at $T = 0.8$ is quite similar in these two cases (with or without consideration of elasticity).

Figure~\ref{fig_circular_compare}g compares the evolution of the domain area for interfaces driven by pure capillarity and the combination of capillarity and elasticity at different temperatures. 
The most obvious effect of raising the temperature is to increase the slope (following the initial transient).
This is largely the result of increasing mobility with increasing temperature. 
Comparing the migration velocities for interfaces with and without disconnection dislocation character shows that  dislocation character (elasticity) slows/enhances the migration rate and that this effect is most significant at low/high temperatures.

Examination of Fig.~\ref{fig_circular_compare}h demonstrates that the degree of faceting/shape anisotropy is larger for interfaces driven by a combination of capillarity and elasticity than those driven by capillarity alone at elevated temperatures; this effect appears to be largest at intermediate temperatures.
At low temperatures, interfaces are strongly faceted under pure capillarity such that adding elastic effects does not significantly alter the faceting. 
At $T = 0.8$, the effective shear-coupling factor $\beta$ is very small (because multiple disconnection modes operate on each reference interface) such that, again, elasticity effects are negligible.
At intermediate temperatures (e.g., $T = 0.1$, $0.3$ or $0.5$), the degrees of faceting and elasticity effects are large.
In all cases, the degree of faceting $\Delta \Phi$  gradually decreases with increasing temperature, but no sharp faceting-defaceting transition is observed.
However, a  large drop in $\Delta \Phi$ is observed for $0.1\leq T \leq 0.3$ for interfaces driven by both capillarity and elasticity, indicating that a sharp faceting-defaceting transition occurs in this temperature range; the sharp transition can also be seen in Fig.~\ref{fig_circular_compare}f. 

\begin{figure*}[t]
\includegraphics[width=1\linewidth]{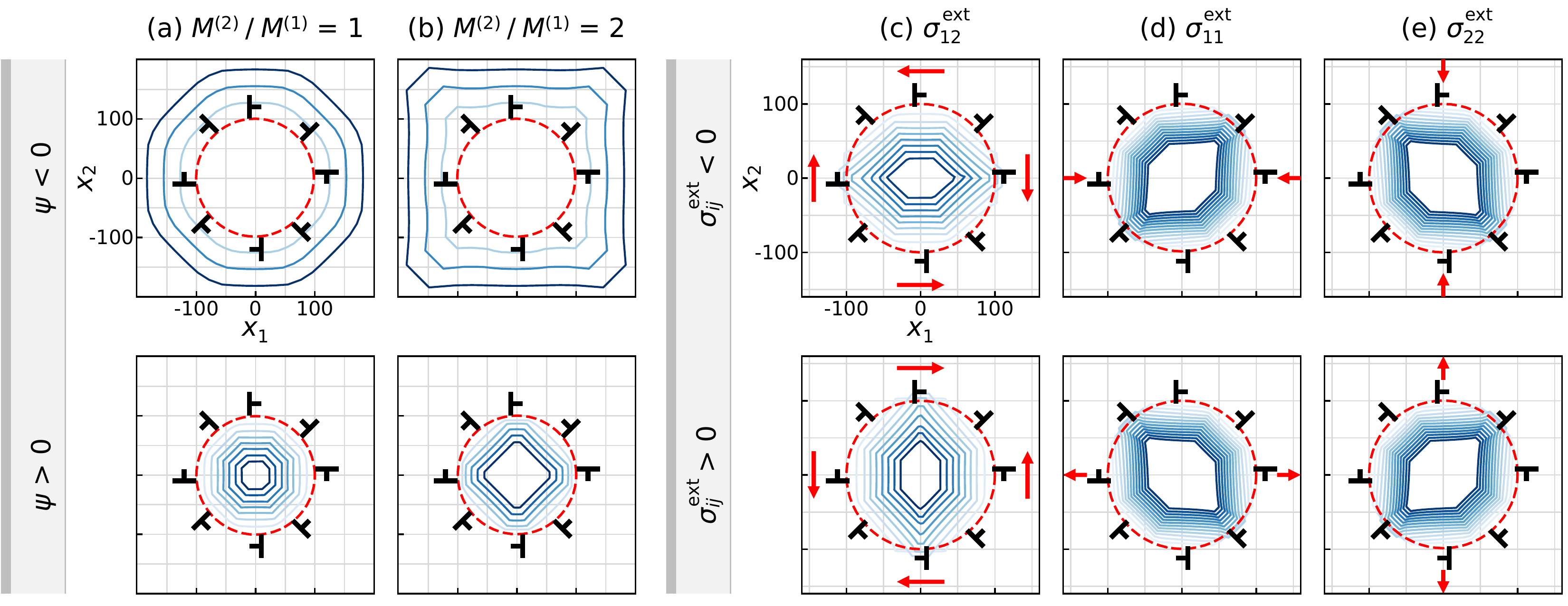}\hspace{-1.78em}%
\caption{Evolution of an initially circular 4-reference interface under different external driving forces. (a), (b) Applying positive and negative chemical potential jumps with different intrinsic mobility anisotropy. 
The red dash lines are the initially circular interface. 
The time interval $\Delta t = 400$. 
(c)-(e) Applying different external stresses  with isotropic intrinsic mobility. }
\label{fig_circular_external}
\end{figure*}

In many situations, some operative disconnection modes may correspond to pure steps according to the dichromatic pattern; e.g., in Fig.~\ref{fig_bicrystallography}d, the disconnections on the reference interface indicated by the green dash line are pure steps.
Figure~\ref{fig_circular_compare}d shows an example where only R(1) and R(3) disconnection possess Burgers vectors while the other two are pure steps (the definition of ``R($k$)'' is shown in Fig.~\ref{fig_bicrystallography}a for the 4-reference pattern'').
In this case, we see a transition of the  shape from an octagon to nearly square as the domain shrinks.  
This is in contrast to the pure capillarity case, Fig.~\ref{fig_circular_compare}a, and the case where all disconnection modes possess finite Burgers vectors, Fig.~\ref{fig_circular_compare}c.
Therefore, the grain shape is strongly influenced by the operative disconnection modes on  different reference interfaces. 
This may explain why the  morphologies, observed in experiments and simulations, are often inconsistent with those predicted based solely upon conventional capillarity~\cite{2012_Radetic_squareAu_exp, 2016_Molodov_circular_MD, 2014_Molodov_circular_MD}.

Faceted domain shape is not only dependent upon intrinsic interface properties (e.g.,  interface energy, elastic interactions between disconnections) but is also influenced by the nature of the driving force (e.g., externally applied stress,  variations in chemical potential). 
The chemical potential may be discontinuous across an interface; e.g., when the materials on the two sides of the interface are of different phases and one of the phases is energetically favorable.
Figure~\ref{fig_circular_external} shows the effects of chemical potential jump and external stress on the evolution starting from the same initial interface morphology at $T = 0.05$. 
The disconnections have step and dislocation character (all other parameters are as in Fig.~\ref{fig_circular_compare}c).

Figures~\ref{fig_circular_external}a and b show the effect of a chemical-potential jump ($\psi$) on domain shape evolution.
The interface has four reference planes (see Fig.~\ref{fig_bicrystallography}a). 
For $\psi < 0$, the chemical potential is lower inside the domain than outside such that the embedded grain grows; when the sign of $\psi$ switches, the embedded grain shrinks.
When the intrinsic mobilities of the four reference interfaces are identical (Fig.~\ref{fig_circular_external}a), the domain shapes during growth and shrinkage are the same (although  corners are sharper for shrinkage).
However, when the intrinsic mobilities are different: $M^{(2)} = M^{(4)} = 2M^{(1)} = 2M^{(3)}$, the embedded domain morphology is completely different upon switching the sign of $\psi$; in particular the overall shape is rotated by $\pi/4$. 
The role of the direction of growth on the final shape may be deduced from the surface normal dependence of the interface velocity (i.e., from the kinetic Wulff shape rather than the Wulff shape~\cite{2005_Du_vplot_theo}). 

Since interface morphology evolves via the motion of disconnections, morphology evolution can be driven by an applied stress. 
Such stress-driven morphology evolution was indeed widely observed in both experiments and simulations~\cite{2009_Monpiou_shearcoupling_exp, 2012_Trautt_shearcoupling_MDPFC,2005_Hanyu_stressfacet_exp}.
In Figs.~\ref{fig_circular_external}c-e, we apply a uniform external stress $\sigma_{12}^{\rm ext}$, $\sigma_{11}^{\rm ext}$, or $\sigma_{22}^{\rm ext}$; the intrinsic mobilities of the four reference interfaces are identical.
Comparing these results without an external stress (i.e. Fig.~\ref{fig_circular_compare}c), we see that the application of a stress leads to the elongation of the grain along one direction and shortening along the perpendicular direction.  
Analysis of the Peach-Koehler (PK) force on each Burgers vector shows that $\sigma_{12}^{\rm ext}$ acts only on R(1) and R(3) disconnections.
The elongation and shortening directions are parallel to the R(1) and R(3) reference interfaces.
A similar analysis applies to the $\sigma_{11}^{\rm ext}$ and  $\sigma_{22}^{\rm ext}$.
In SM, we demonstrate  what applied stress conditions lead to grain growth/shrinkage.

Our analysis of a single, embedded domain shows that disconnection dislocation character can produce marked changes in domain morphology evolution as compared with pure steps/capillarity-driven dynamics. 
These effects are associated with disconnection stress fields,  elastic interactions between disconnections, and their interaction with external stresses. 
The results are consistent with experiments and show how the morphology may be manipulated by applying stress during annealing.

\section{Steady-State Interface Migration: embedded half-loops}

An intrinsic difficulty in the experimental determination of interface mobility is that the capillarity driving force typically decreases during microstructural evolution (e.g., in normal grain growth) or increases during shrinkage of an embedded grain (Fig.~\ref{configurations}a). 
Such changes modify the relative importances of the different driving forces. 
Shvindlerman, Gottstein, and co-workers~\cite{halfloop_book, 1973_Aristov_halfloop_exp, 2008_Kirch_facetcapillarity_halfloop_exp} introduced a half-loop bicrystal geometry, as shown in Fig.~\ref{configurations}b, for which the capillarity driving force is constant, leading to steady-state GB migration.
This has been exploited in a wide range of experimental~\cite{2008_Straumal_Zinc1_exp, 2008_Sursaeva_Zinc2_exp,2008_Kirch_facetcapillarity_halfloop_exp} and atomistic simulation~\cite{1998_Upmanyu_halfloopAl_MD} studies~\cite{halfloop_book}. 
We exploit this geometry to observe the influence of elastic interactions on interface faceting-defaceting transitions and migration.

\begin{figure}[t]
\includegraphics[width=1\linewidth]{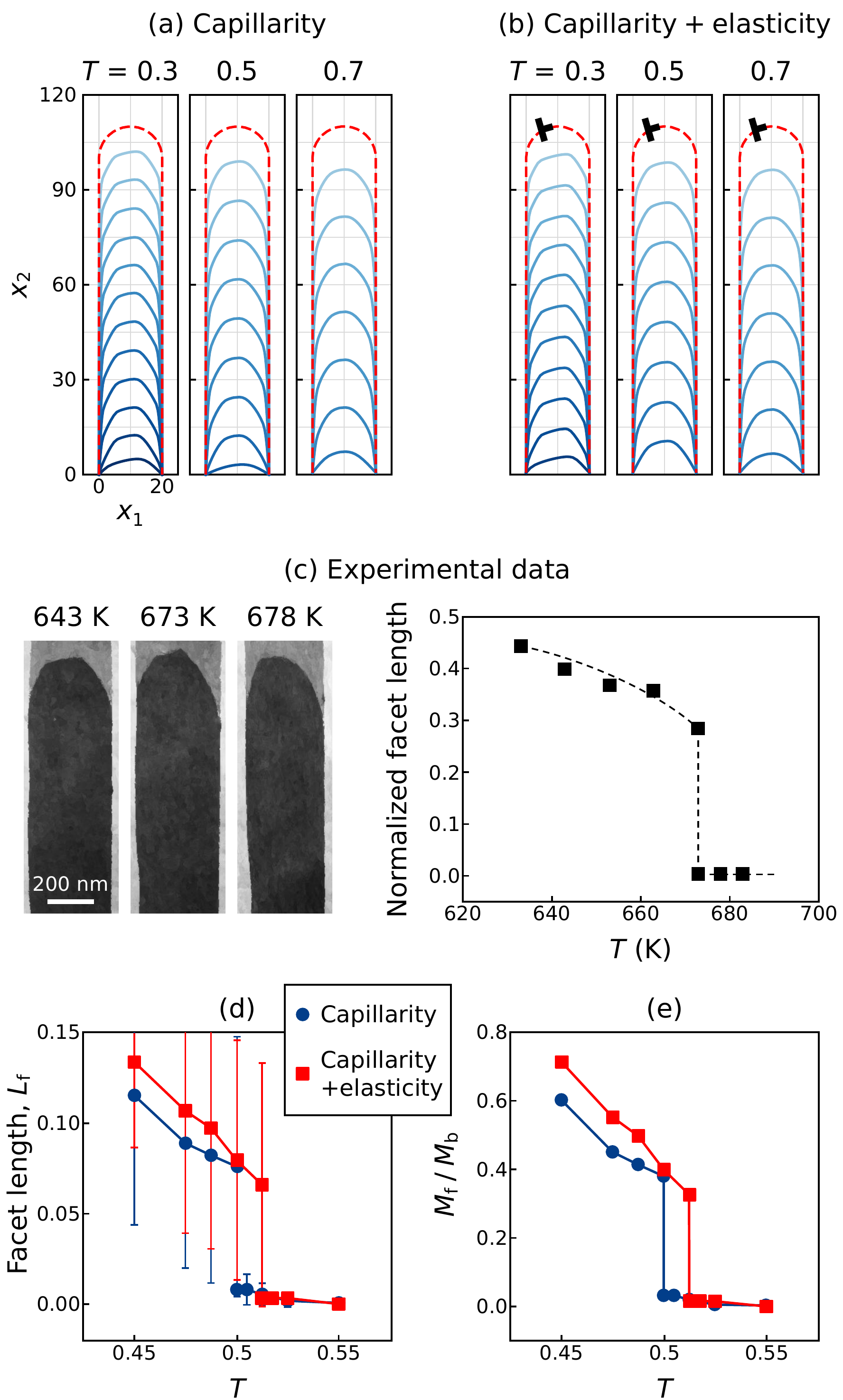}\hspace{-1.78em}%
\caption{Numerical results for the evolution of a $[10\bar{1}0]$ tilt GB half-loop in Zn driven by (a) pure capillarity and (b) a combination of capillarity and elasticity at different temperatures. 
The red dash lines are the initial half-loop GBs. 
The time interval is $\Delta t = 50$. 
(c) Experimental morphologies of the $[10\bar{1}0]$ tilt GB half-loop in Zn and the corresponding normalized facet length vs. temperature, reproduced with permission from~\cite{2008_Straumal_Zinc1_exp,2008_Sursaeva_Zinc2_exp} (Copyright 2008 Taylor $\&$ Francis; 2008 Elsevier). 
(d) Facet length $L_{\rm f}$ vs. temperature corresponding to  (a) (blue  circles) and (b) (red squares). 
(e) Normalized facet mobility $M_{\rm f}/M_{\rm b}$ vs. temperature obtained by Eq.~\eqref{MfMb}. 
(d) and (e) share the legend. 
} 
\label{fig_halfloop_Zn}
\end{figure}

Figure~\ref{fig_halfloop_Zn} shows the evolution of a half-loop grain boundary in a hexagonal close-packed Zn bicrystal that undergoes a faceting-defaceting transition with temperature.
The CSL of the $[10\bar{1}0]$ tilt misoriented crystals corresponds to  2-reference interfaces; the (0001) close-packed plane and the $\sim 76^\circ$ inclined from the (0001) plane~\cite{2008_Straumal_Zinc1_exp}. 
Referring to Figs.~\ref{fig_halfloop_Zn}a-c, the (0001) plane is parallel to the long edges of the half-loop. 
The experiments suggest that the activation energy for the motion of the $\sim 76^\circ$ facets (0.1 eV) is much smaller than that for the overall shrinkage rate of the half loop (1-2 eV).
Therefore, we set the activation energy for the motion of disconnections on this facet to be 10\% that of those on the (0001) reference interface.

Two cases are investigated: (i) pure step disconnections that are  only  driven by capillarity, and (ii) the disconnections with step and dislocation character, where both capillarity and elasticity contribute to the driving force. 
The results are shown in Figs.~\ref{fig_halfloop_Zn}a and b.
In both cases, a facet forms near the top of the half-loop at $T = 0.3$, and the angle between the facet and the vertical (0001) reference interface is 76$^\circ$, in agreement with our CSL analysis. 
The facet length formed in case (ii) is larger than in case (i) while in both cases the facets are significantly shortened at $T=0.5$, and completely disappear by $T = 0.7$; the top of the half-loop is smoothly curved. 
Such a faceting-defaceting transition was also observed in the experiments, where the facet length decreased upon raising the annealing temperature from 643 K to 673 K, and a defaceted morphology was observed at 678 K (see Fig.~\ref{fig_halfloop_Zn}c~\cite{2008_Straumal_Zinc1_exp}).

Figure~\ref{fig_halfloop_Zn}d shows the relation between the steady-state facet length $L_\rmf$ and temperature.
$L_\rmf$ gradually decreases with increasing temperature and then drops abruptly to zero (i.e., the loop becomes smooth) at $T =0.510$ for pure capillarity or at $T = 0.525$ when elasticity is considered; this suggests that  dislocation character enhances  interface facet stability. 
The first-order nature of the phase transition is consistent with the experimental observations in Fig.~\ref{fig_halfloop_Zn}c~\cite{2008_Sursaeva_Zinc2_exp}.

GB mobility cannot be directly measured from the evolution of the half-loop. 
Based on a pure capillarity assumption, Sursaeva, et al.~\cite{2008_Sursaeva_Zinc2_exp} derived an expression for the normalized facet mobility (the ratio between  facet mobility and  total GB mobility):
\begin{equation}\label{MfMb}
\frac{M_\rmf}{M_\rmb} = \frac{\theta - \phi}{\sin \theta \sin \phi} \frac{2L_\rmf}{a-2L_\rmf\sin\theta}, 
\end{equation}
where $\theta$ is the misorientation angle of the $[10\bar{1}0]$ tilt GB, $\phi$ is the inclination angle of the facet, and $a$ is the half-loop width. 
Applying this expression to our half-loops for disconnections with and without dislocation character, we find that the normalized facet mobility is $\sim20\%$ larger upon including elastic effects than under pure capillarity at $T <  0.5$; see Fig.~\ref{fig_halfloop_Zn}e.
A first-order transition in the  facet mobility also occurs at the temperature where the abrupt decrease in facet length is observed.

\begin{figure}[t]
\includegraphics[width=1\linewidth]{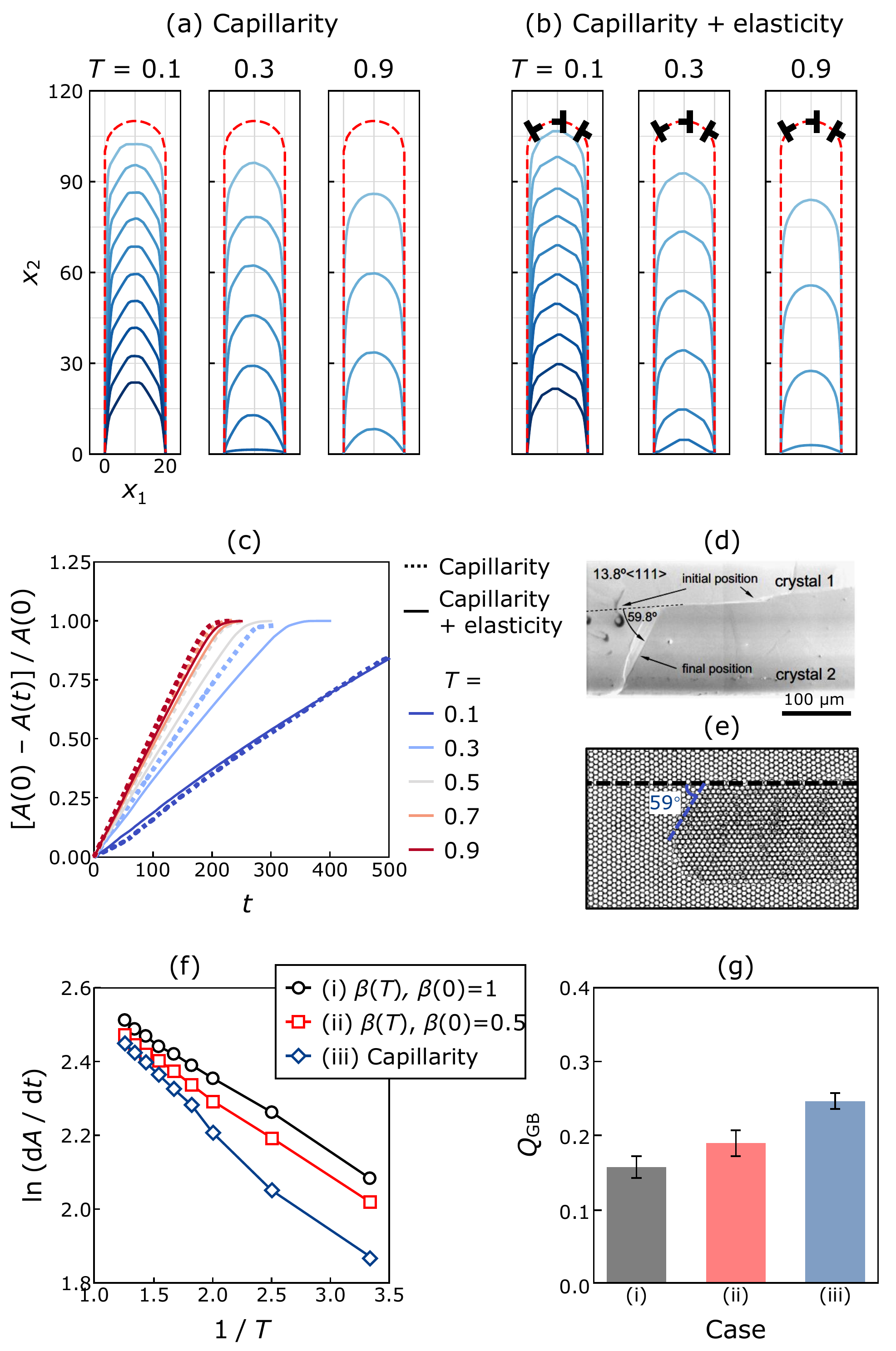}\hspace{-1.78em}%
\caption{Numerical results for the evolution of a $[111]$ tilt GB half-loop driven by (a) pure capillarity and (b) a combination of capillarity and elasticity at several temperatures
The red dash lines represent the initial half-loops. 
(c) The normalized area $[A(0)-A(t)]/A(0)$ vs. time of the $[111]$ tilt GB half-loop. (d) Experimental observations and (e) atomistic simulation results of the $[111]$ tilt GB quarter-(or half-) loop in Al,  reproduced with permission from~\cite{2008_Kirch_facetcapillarity_halfloop_exp, 1998_Upmanyu_halfloopAl_MD} (Copyright 2008 Elsevier; 1998 Springer).
(e) Reduced migration mobility $\ln(\rmd A/ \rmd t)$ vs. $1/T$ and (d) the corresponding apparent activation energy of the half-loop shrinkage driven by pure capillarity and elasticity with different shear-coupling factors. 
(d) and (e) share the legend. 
}
\label{fig_halfloop_Al}
\end{figure}

Evolution of a quarter-loop (Fig.~\ref{configurations}b) of $[111]$ tilt GB in Al was also  studied experimentally~\cite{2008_Kirch_facetcapillarity_halfloop_exp}.
We performed numerical simulations on the same GB in Al (but with the half-loop geometry)  to deduce the effect of elastic interactions (see  Fig.~\ref{fig_halfloop_Al}).
Figures~\ref{fig_halfloop_Al}a and b show that the GBs are faceted at low temperature and fully defaceted at high temperature; 
the faceting-defaceting transition temperature is $T \simeq 0.5$.
The FCC $[111]$ tilt dichromatic pattern in Fig.~\ref{fig_bicrystallography}c implies this is a 6-reference case. 
We chose the R(4) interface ($\phi^{(4)} = 90^\circ$) as the two edges of the half-loop such that possible facet angles are $\phi = 0^\circ$, $30^\circ$ and $60^\circ$ relative to the $\mathbf{e}_1$-axis.
The GB profile shows $0^\circ$ and $60^\circ$  facets without elasticity (Fig.~\ref{fig_halfloop_Al}a) and only $30^\circ$ when elasticity (dislocation character) is included (Fig.~\ref{fig_halfloop_Al}b).
Both experimental observations and simulation results (see Fig.~\ref{fig_halfloop_Al}d, e) only show  $30^\circ$ facets ($60 ^\circ$ away from the long edge)~\cite{2008_Kirch_facetcapillarity_halfloop_exp, 1998_Upmanyu_halfloopAl_MD}, consistent with the numerical results which include dislocation character  (Fig.~\ref{fig_halfloop_Al}b).

Figure~\ref{fig_halfloop_Al}c shows  numerical results for the temporal evolution of the half-loop area at several temperatures.
For $T>0.1$, the GB migration velocities evolving under the combined effects of capillarity and elasticity are larger than those driven by capillarity alone. 
GB migration velocities and activation energies are shown in Fig.~\ref{fig_halfloop_Al}f.
The blue and black curves correspond to the cases shown in Fig.~\ref{fig_halfloop_Al}a (pure capillarity) and b (both capillarity and elasticity - effective shear-coupling factors $\beta=1$ at $T= 0$).
We also consider a similar case with $\beta=0.5$ at $T= 0$ to study the effect of  $\beta$ on GB migration. 
At low temperatures ($T \leq 0.5$), the migration velocities of GBs driven by combined capillarity and elasticity are much larger than those driven by pure capillarity. 
With temperature increasing, GB migration velocities for cases including elasticity approach those without elasticity; this is expected since the effective shear-coupling factor $\beta\to 0$ with increasing temperature. 
The apparent activation energies ($Q_{\rm GB}$) shown in the bar chart in Fig.~\ref{fig_halfloop_Al}g are deduced from the slope of the curves in Fig.~\ref{fig_halfloop_Al}f for $T \geq 0.5$. 
Apparent activation energies for $\beta(T)$ case where $\beta(0) = 1$ and $\beta(0) = 0.5$ are 0.16 and 0.19, which are  48.4$\%$ and 38.7$\%$ smaller than that for the pure capillarity case ($Q_{\rm GB} = 0.25$).
This clearly demonstrate that the elastic fields disconnection Burgers vectors strongly affect the apparent activation energies for GB migration.

\section{Nominally flat interfaces}

\begin{figure*}[t]
\includegraphics[width=0.96\linewidth]{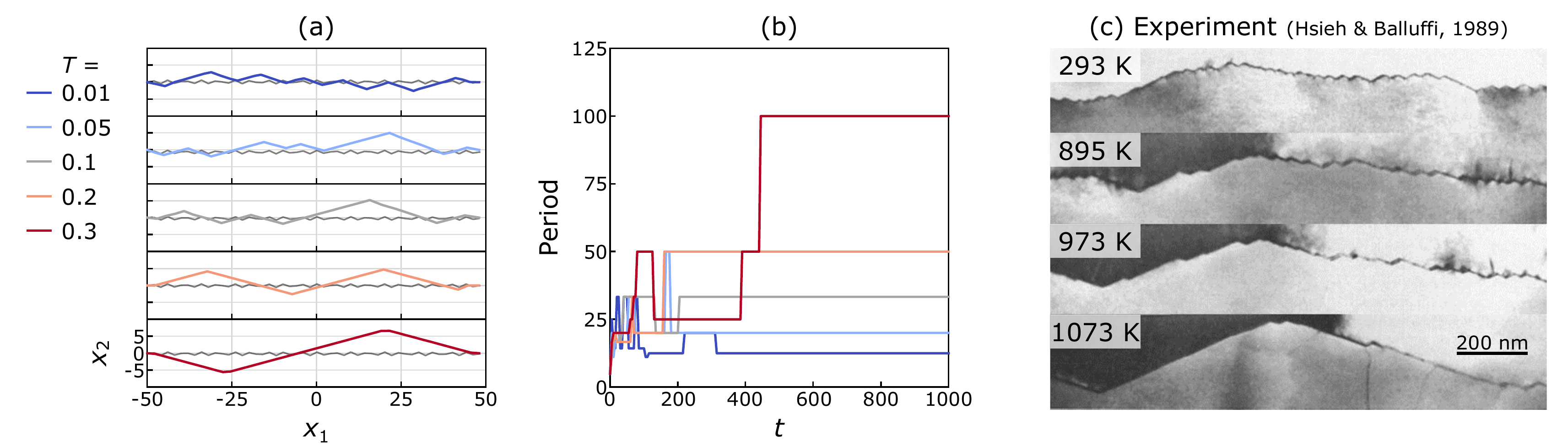}\hspace{-1.78em}%
\caption{Evolution of an initially unfaceted (microscopically rough) flat interface with 6 references under fixed time annealing. (a) The profile of an initially unfaceted flat interface (dark gray lines)  and the interface profile at $t=1000$ at different temperatures. 
(b) Facet period vs. time for the interfaces shown in (a). 
(a) and (b) share the legend. 
(c) Experimental morphologies of a faceted $[111]$ tilt GB at several temperatures. 
At higher temperature, larger degree of facet coarsening was observed. 
The images are reproduced with permission from~\cite{1989_Balluffi_defaceting_exp} (Copyright 1989 Elsevier).  }
\label{fig_flat_coarsening}
\end{figure*}

Since  interfaces  evolve in the embedded grain and GB half-loop configurations, deducing the nature of the thermodynamic faceting-defaceting transition is difficult.
We now consider the case of a nominally flat interface pinned at two ends, as schematically shown in Fig.~\ref{configurations}c; this can be viewed as a simplified model for a GB in a polycrystalline microstructure where GBs are delimited by  triple junctions (TJ). 
GB motion is highly constrained by the low mobility of the TJs. 
This model is reasonable since faceting-defaceting behavior is dominated by the intrinsic behavior of the GBs and the effects of  TJs on faceting is  weak. 
Therefore, we make the approximation that TJs are immobile (i.e., interfaces are pinned at TJs) for this  faceting-defaceting study.

The interface pinned at its ends, where the vector between the two pinning points is not a facet direction is interesting.
In this case, in the absence of elasticity, the energy of the fully faceted interface (anisotropic interface energy) is independent of the faceting amplitude and period; the length of this interface as the Manhattan distance (see SM).
This implies, in the anisotropic interface energy case, there is no faceting/defaceting transition  \cite{hausser2005facet,hamilton2006}.
Hence, athermodynamic faceting-defaceting transition must include elastic effects; i.e., the  long-range elastic interaction between disconnections with dislocation character.
In this section, we focus on the thermodynamics and kinetics of nominally flat, pinned interfaces.

Figure~\ref{fig_flat_coarsening}a shows the temporal evolution of an initially unfaceted, nominally flat (microscopically rough) interface with 6 references and inclination $\phi=45^\circ$ at several temperatures. 
Referring to the example 6-reference pattern in Fig.~\ref{fig_bicrystallography}c, the $\phi = 45^\circ$ interface can be described by  $\phi=30^\circ$ and $60^\circ$ reference interfaces (i.e., R(2) and R(3) interfaces).
As shown in Fig.~\ref{fig_flat_coarsening}a, the initially flat interface (dark gray lines) facets on progressively larger (amplitude and period) scales with increasing temperature (fixed time annealing).  
The facets that appear on annealing exhibit consistent largely of two inclinations and meet at facet junctions  at turning angles of $\pm30^\circ$ ($\pm15^\circ$ with respect to the  flat interface).
The faceted interface profiles are in excellent agreement with those experimentally observed for $[111]$ tilt GBs in Au (Fig.~\ref{fig_flat_coarsening}c)~\cite{1989_Balluffi_defaceting_exp}.

Continuum elasticity analysis and atomistic simulations~\cite{2003_Hamilton_stresseffect_theo, 2009_Wu_stresseffect_MD} show that the elastic energy decreases with increasing facet period; in other words, the faceted interface with the largest possible facets is thermodynamically preferred to the flat interface at low temperature.
At the same time, the presence of finer facets at low temperatures in isochronal annealing treatments (numerical results and experiments) suggests that facet coarsening is kinetically controlled. 
Figure~\ref{fig_flat_coarsening}b shows the temporal evolution of the period of the interface profile under different temperatures (period  obtained by  Fourier analysis of the interface profile). 
At equilibrium (i.e., after very long time anneals), the interface should be faceted with a period 100 (i.e., the distance between two pinned ends). 
At  low temperature ($T=0.01$), facet coarsening is too slow to the achieve equilibrium interface facet profile within the explored simulation times. 
At  high temperature ($T=0.3$), the faceted profile is fully coarsened within the simulation time. 
The coarsening process is illustrated in Fig.~\ref{fig_flat_coarsening}b. 
Each jump corresponds to the merging of two neighboring facet junctions.

\begin{figure*}[t]
\includegraphics[width=1\linewidth]{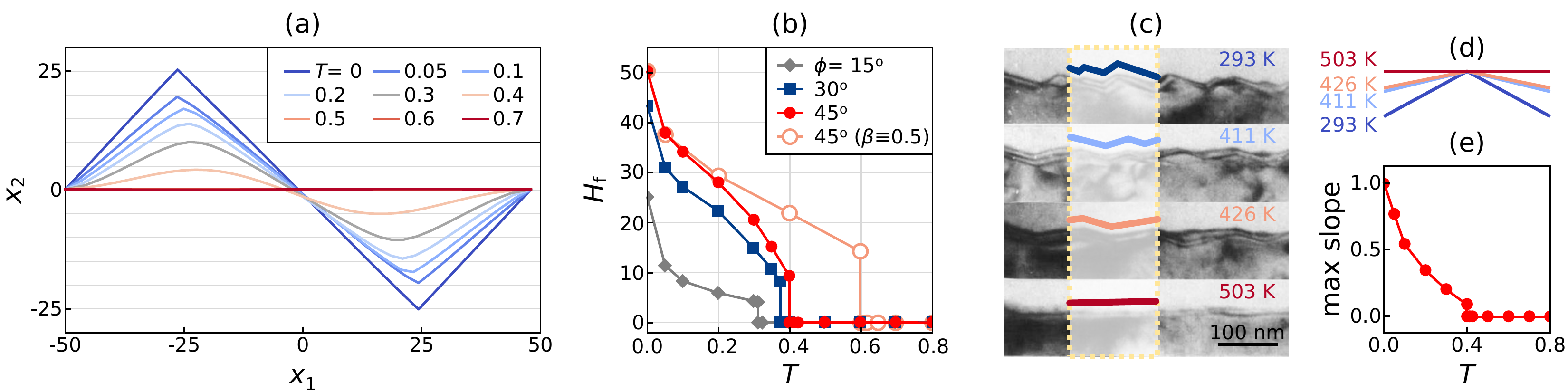}\hspace{-1.78em}%
\caption{Thermodynamic faceting-defaceting transition of a flat interface with 2 references.
(a) Equilibrium morphologies of a flat interface at different temperatures. 
(b) Facet heights of the interface profiles with different inclinations vs. temperature. 
(c) Experimental observations of facet-defaceting transition of $[111]$ tilt GBs in Al~\cite{1989_Balluffi_defaceting_exp}; reproduced with permission (Copyright 1989 Elsevier). 
(d) The interface slopes at different temperatures extracted from the experimental images in (c).
(e) Maximum slope of the interface profile at different temperatures for $\phi = 45^\circ$.  
}
\label{fig_flat_defaceting}
\end{figure*}

With continued increases in temperature, the interface may undergo a faceting-defaceting transition. 
To study this phenomenon, we examine the equilibrium profile of a 2-reference interface (e.g., Fig.~\ref{fig_bicrystallography}b) with inclination $\phi= 45^\circ$ at several temperatures. 
We initialize the simulations with a fully-coarsened faceted interface (equilibrium or near-equilibrium configuration below the transition temperature). 
The simulations are performed for times sufficient to achieve equilibrium interface profiles (i.e., until profile evolution ceases). 
Figure~\ref{fig_flat_defaceting}a shows the equilibrium interface profiles at different temperatures. 
We see that the interface profile amplitude $H_\rmf$ (defined as the difference in height between the peak and the valley) gradually decreases with increasing temperature; the interface undergoes a faceting-defaceting transition  ($H_\rmf = 0$) at $0.4<T<0.5$.
Figure~\ref{fig_flat_defaceting}b shows $H_\rmf$ vs. temperature. 
The red curve in Fig.~\ref{fig_flat_defaceting}b corresponds to the case of Fig.~\ref{fig_flat_defaceting}a. 
$H_\rmf$ abruptly drops from 9.34 to near 0 at $T = 0.41$; this  indicates that the faceting-defaceting transition is first-order.
Such first-order faceting-defacting transitions have been observed   experimentally; e.g., the $[111]$ tilt GBs in Al~\cite{1989_Balluffi_defaceting_exp}, as seen in Fig.~\ref{fig_flat_defaceting}c.

The elastic energy dominates the low-temperature behavior and favors a faceted profile (driving disconnections toward the facet junctions), while the configurational entropy dominates the high-temperature behavior and favors an unfaceted profile (i.e., a uniform spatial distribution of disconnections); this leads to a  first-order facet-defaceting transition  as described further  in SM. 
The effective shear-coupling factor $\beta$ is also temperature-dependent. 
This dependence does not change the nature of the first-order transition, but does lower the transition temperature; cf. the case where $\beta$ is temperature dependent and that where $\beta$ is temperature independent (solid red circles and the unfilled circles in Fig.~\ref{fig_flat_defaceting}b).  
Since $\beta\to0$ as temperature increases, including the temperature dependence of $\beta$ further decreases the elastic energy with temperature increasing. 

It is interesting to note that, as shown in Fig.~\ref{fig_flat_defaceting}e, increasing temperature(below the transition temperature), the maximum slopes of the equilibrium profiles increasingly deviate from the reference inclinations (slopes $\pm1$). 
This phenomenon is also seen experimentally~\cite{1989_Balluffi_defaceting_exp}. 
We extracted the interface slopes from the experimental images in Fig.~\ref{fig_flat_defaceting}c at different temperatures in Fig.~\ref{fig_flat_defaceting}d;  the maximum slope decreases with increasing  temperature. 
 
Fig.~\ref{fig_flat_defaceting}b shows that the faceting-defaceting transition temperature decreases with decreasing interface inclination ($\phi = 45^\circ$, $30^\circ$ and $15^\circ$). Decreasing $\phi$ implies decreasing disconnection density.


\section{Conclusions}

In this study, we examined the faceting and defaceting of internal interfaces in crystalline materials based upon a mechanistic model that accounts for bicrystallography, interface structure, and elasticity, in addition to capillarity. 
We demonstrate that many of the experimentally-observed features of the faceting-defaceting transition and  interface morphology evolution may be traced to the thermodynamics and dynamics of the disconnections that lie in the interface.  
Disconnections are line defects with  step and/or dislocation character determined by bicrystallography.
The latter implies that elasticity plays an important role in the properties/behavior of internal interfaces. 
Compared with conventional theories for interface motion, based upon anisotropic surface energy, the elastic effects associated with disconnections  
\begin{itemize}
\item
lead to a thermodynamic, first-order, finite-temperature faceting-defaceting transition,
\item
enhance faceting during microstructure evolution and profoundly modify interface morphologies,
\item
make interface migration/morphology-evolution stress-dependent, and
\item
increase the apparent mobility of an entire interface whilst reducing the apparent activation energy for interface migration.
\end{itemize}
We demonstrate  these phenomena through numerical simulations based upon a general, rigorous, continuum disconnection-based model for interface (grain boundaries, heterophase interfaces) thermodynamics and kinetics~\cite{2022_Han_EOM_theory}. 
Our simulations examine embedded particles/grains, steady-state interface migration geometries (half-loops~\cite{1973_Aristov_halfloop_exp}), and nominally flat interfaces pinned between junctions.
The corresponding results demonstrate that consideration of the elastic fields associated with disconnections is essential for describing both equilibrium morphologies and morphology/microstructure evolution in crystalline systems with internal interfaces.

\section{Methods}

The equation of motion for disconnection-mediated interface migration  (see~\cite{2017_Zhang_EOM_theory,2022_Han_EOM_theory}) is Eq.~\eqref{EOM}. 
A more detailed account of this continuum model and the expressions for each term in Eq.~\eqref{EOM} are provided in SM.
The simulations are performed by solving Eq.~\eqref{EOM} by a finite difference method. 
The interface  is discretized into $N$ segments, the coordinates of these points are $\mathbf{x}_i$ with $i = 0,1,\dots, N-1$. 
Structure-related quantities (e.g., curvature, local tangent, $\ldots$) are calculated based on derivatives along the interface curve using a three-point stencil.

For all the simulations,  reduced quantities are employed: $\tilde{\mathbf{x}} = \mathbf{x}/\alpha$ (for all ``length'' quantities; $\alpha$ is the length of one DSC cell edge), the time interval $\Delta\tilde{t} = \Delta t M^{(1)}\gamma^{(1)}/\alpha^2$, the intrinsic mobility $\tilde{M}^{(k)} = M^{(k)}/M^{(1)}$, the stress $\tilde{\tau} = \tau\alpha/\gamma^{(1)}$, the chemical potential jump $\tilde{\psi} = \psi\alpha/\gamma^{(1)}$, the temperature $\tilde{T} = k_\rmB T/(\gamma^{(1)}\alpha^2)$ ($k_\rmB$ is the Boltzmann constant), the activation energy $\tilde{Q} =Q/(\gamma^{(1)}\alpha^2)$, and the interface energy $\tilde{\gamma} =\gamma/\gamma^{(1)}$. 
Throughout the paper, we omit the tilde symbol for simplicity. 

The initial shape of the embedded grain is a circle with a radius of 100. 
The width and height of the initial GB half-loop are set as 20 and 110, respectively.
The length and inclination angle of the initially unfaceted, pinned interface is 100 and $45^\circ$, respectively.

\begin{table*}[tb]
\caption{\label{tab:parameters}Parameters used in the simulations. 
``$e$'' is the base of natural logarithm.
$m$ is odd. }
\renewcommand{\arraystretch}{1.5}
\begin{ruledtabular}
\begin{tabular}{ccccccccccccc}
\multirow{2}{*}{} & 
\multirow{2}{*}{Figure No.} & 
\multirow{2}{*}{$n$-reference} & 
\multicolumn{2}{c}{Interface energy} & 
\multicolumn{4}{c}{Shear-coupling factor} & 
\multicolumn{4}{c}{Intrinsic mobility} \\ 
\cline{4-5} \cline{6-9} \cline{10-13}
 & & & 
$\gamma^{(m)}$ & $\gamma^{(m+1)}$ & 
$\beta_1$ & $E^*_1$ & $\beta_2$ & $E^*_2$ & 
$M_0^{(m)}$ & $M_0^{(m+1)}$ & $Q^{(m)}$ & $Q^{(m+1)}$ \\ 
\hline
\multirow{5}{*}{\shortstack{Embedded\\grain}} & 
\ref{fig_circular_equ}b, c, d & 
2, 4, 6 & 
1 & 1 & 
0 & - & 0 & - & 
1 & 1 & - & - \\ 
 & 
\ref{fig_circular_compare}a, b, e & 
4 & 
1 & 1 & 
0 & - & 0 & - & 
$e$ & $e$ & 0.01 & 0.01 \\ 
 & 
\ref{fig_circular_compare}c, d, f & 
4 & 
1 & 1 & 
0.5 & 1 & $-0.25$ & 2 & 
$e$ & $e$ & 0.01 & 0.01 \\ 
 & \ref{fig_circular_external}a, c, d, e & 
4 &
1 & 1 & 
0.5 & 1 & $-0.25$ & 2 & 
$e$ & $e$ & 0.01 & 0.01 \\ 
 & 
\ref{fig_circular_external}b & 
4 &  
1 & 1 & 
0.5 & 1 & $-0.25$ & 2 & 
$0.5e$ & $e$ & 0.01 & 0.01 \\ 
\hline
\multirow{4}{*}{\shortstack{GB\\half-loop}} & 
\ref{fig_halfloop_Zn}a (Zn) & 
2 &
1 & 1 & 
0 & - & 0 & - & 
$e$ & $0.5e$ & 0.01 & 0.1 \\ 
 & 
\ref{fig_halfloop_Zn}b (Zn) &
2 &
1 & 1 & 
1 & 1 & $-0.5$ & 2 & 
$e$ & $0.5e$ & 0.01 & 0.1 \\ 
 & 
\ref{fig_halfloop_Al}a (Al) & 
6 & 
1 & 1 & 
0 & - & 0 & - & 
$e$ & $e$ & 0.01 & 0.01 \\ 
 & 
\ref{fig_halfloop_Al}b (Al) & 
6 &
1 & 1 & 
1 & 1 & $-0.5$ & 2 & 
$e$ & $e$ & 0.01 & 0.01 \\ 
\hline
\multirow{2}{*}{\shortstack{Flat\\interface}} & 
\ref{fig_flat_coarsening} & 
6 & 
1 & 1 & 
0.5 & 1 & $-0.25$ & 2 & 
$e$ & $e$ & 0.01 & 0.01 \\ 
 & 
\ref{fig_flat_defaceting} &
2 &
1 & 1 & 
0.5 & 1 & $-0.25$ & 2 & 
$e$ & $e$ & 0.01 & 0.01 
\end{tabular}
\end{ruledtabular}
\end{table*}

Based on a solid-on-solid approximation, the temperature dependence of the interface free energy can be described by~\cite{1951_Burton_surfaceenergy_theory,1982_Avron_SOS_theory}:  
\begin{widetext}
\begin{align}
\gamma(\phi,T)
&= \gamma(\phi,0)
+ k_\rmB T \left[
|\sin\phi| \ln\left(\frac{|\sin\phi|}{|\sin\phi| + |\cos\phi|}\right)
+ |\cos\phi| \ln\left(\frac{|\cos\phi|}{|\sin\phi| + |\cos\phi|}\right)
\right]
\nonumber\\
&+ k_\rmB T \left[
|\sin\phi| \ln\left(\frac{1+\epsilon+R|\cot\phi|}{2}\right)
+ |\cos\phi| \ln\left(\frac{1+\epsilon-R}{(1-\epsilon)(1-|\tan\phi|)}\right)
\right],
\end{align}
\end{widetext}
where the zero-temperature interface energy $\gamma(\phi,0)$ is Eq.~\eqref{interface_energy_phi}, $\epsilon = e^{-4\gamma^{(1)}/k_\rmB T}$ and $R=\sqrt{(1-\epsilon)^2\tan^2\phi + 4\epsilon}$. 
The parameters required for $\gamma(\phi,T)$ are $\{\gamma^{(k)}\}$ ($k=1,\dots, n$); see Table~\ref{tab:parameters}. 
The temperature dependence of the effective shear-coupling factor on each reference interface is  obtained by averaging over different disconnection Burgers vectors and step heights weighted by Boltzmann factors~\cite{2019_Chen_disconnection_MC}. 
Here we assume that two disconnection modes are activated. 
The effective shear-coupling factor on R($k$) interface is 
\begin{equation}
\beta^{(k)}(T)
= \frac{\sum_{i=1,2} b_i e^{-E^*_i w/k_\rmB T}}
{\sum_{i=1,2} h_i e^{-E^*_i w/k_\rmB T}}, 
\end{equation}
where $b_i$, $h_i$ and $E^*_i$ ($i=1,2$) are the Burgers vector, the step height and the nucleation barrier for the $i^\text{th}$ disconnection mode, and $w$ is the thickness of the thin film sample. 
The reduced nucleation barrier is defined as $\tilde{E}^* = E^*w/(\gamma^{(1)}\alpha^2)$. 
In some simulations, we set $\beta^{(k)}$ to zero for some reference interfaces. 
The parameters required for $\beta^{(k)}(T)$ are $(\beta_1, E^*_1)$ and $(\beta_2, E^*_2)$; see  Table~\ref{tab:parameters}.  
The temperature dependence of the intrinsic mobility of R($k$) reference interface is assumed to be Arrhenius: 
\begin{equation}
M^{(k)}(T)
= M_0^{(k)} e^{-Q^{(k)}/k_\rmB T}, 
\end{equation}
where $Q^{(k)}$ is the activation energy and $M_0^{(k)}$ is the prefactor; see Table~\ref{tab:parameters}.

\section{Acknowledgements}
JH acknowledges support from City University of Hong Kong Start-up Grant 7200667, the Early Career Scheme (ECS) of Hong Kong RGC Grant 9048213 and Donation for Research Projects 9229061. 
DJS acknowledges support from the Hong Kong Research Grants Council Collaborative Research Fund C1005-19G. 
MS acknowledges the support of the Emmy Noether Programme of the German Research Foundation (DFG) under Grant SA4032/2-1.

\bibliography{mybib}

\begin{thebibliography}{44}%
\makeatletter
\providecommand \@ifxundefined [1]{%
 \@ifx{#1\undefined}
}%
\providecommand \@ifnum [1]{%
 \ifnum #1\expandafter \@firstoftwo
 \else \expandafter \@secondoftwo
 \fi
}%
\providecommand \@ifx [1]{%
 \ifx #1\expandafter \@firstoftwo
 \else \expandafter \@secondoftwo
 \fi
}%
\providecommand \natexlab [1]{#1}%
\providecommand \enquote  [1]{``#1''}%
\providecommand \bibnamefont  [1]{#1}%
\providecommand \bibfnamefont [1]{#1}%
\providecommand \citenamefont [1]{#1}%
\providecommand \href@noop [0]{\@secondoftwo}%
\providecommand \href [0]{\begingroup \@sanitize@url \@href}%
\providecommand \@href[1]{\@@startlink{#1}\@@href}%
\providecommand \@@href[1]{\endgroup#1\@@endlink}%
\providecommand \@sanitize@url [0]{\catcode `\\12\catcode `\$12\catcode
  `\&12\catcode `\#12\catcode `\^12\catcode `\_12\catcode `\%12\relax}%
\providecommand \@@startlink[1]{}%
\providecommand \@@endlink[0]{}%
\providecommand \url  [0]{\begingroup\@sanitize@url \@url }%
\providecommand \@url [1]{\endgroup\@href {#1}{\urlprefix }}%
\providecommand \urlprefix  [0]{URL }%
\providecommand \Eprint [0]{\href }%
\providecommand \doibase [0]{https://doi.org/}%
\providecommand \selectlanguage [0]{\@gobble}%
\providecommand \bibinfo  [0]{\@secondoftwo}%
\providecommand \bibfield  [0]{\@secondoftwo}%
\providecommand \translation [1]{[#1]}%
\providecommand \BibitemOpen [0]{}%
\providecommand \bibitemStop [0]{}%
\providecommand \bibitemNoStop [0]{.\EOS\space}%
\providecommand \EOS [0]{\spacefactor3000\relax}%
\providecommand \BibitemShut  [1]{\csname bibitem#1\endcsname}%
\let\auto@bib@innerbib\@empty
\bibitem [{\citenamefont {Wulff}(1901)}]{1901_Wulff_theory}%
  \BibitemOpen
  \bibfield  {author} {\bibinfo {author} {\bibfnamefont {G.}~\bibnamefont
  {Wulff}},\ }\bibfield  {title} {\bibinfo {title} {On the question of speed of
  growth and dissolution of crystal surfaces},\ }\href
  {https://doi.org/https://doi.org/10.1524/zkri.1901.34.1.449} {\bibfield
  {journal} {\bibinfo  {journal} {Zeitschrift Für Kristallographie -
  Crystalline Materials.}\ }\textbf {\bibinfo {volume} {34}},\ \bibinfo {pages}
  {449– 530} (\bibinfo {year} {1901})}\BibitemShut {NoStop}%
\bibitem [{\citenamefont {Frank}(1958)}]{1958_frank}%
  \BibitemOpen
  \bibfield  {author} {\bibinfo {author} {\bibfnamefont {F.~C.}\ \bibnamefont
  {Frank}},\ }\bibfield  {title} {\bibinfo {title} {{On the Kinematic Theory of
  Crystal Growth and Dissolution Processes}},\ }in\ \href@noop {} {\emph
  {\bibinfo {booktitle} {Growth and Perfection in Crystals}}},\ \bibinfo
  {editor} {edited by\ \bibinfo {editor} {\bibfnamefont {T.~D.}\ \bibnamefont
  {Doremus R.~H.}, \bibfnamefont {Roberts B.~W.}}}\ (\bibinfo  {publisher}
  {John Wiley \& Sons},\ \bibinfo {address} {New York},\ \bibinfo {year}
  {1958})\BibitemShut {NoStop}%
\bibitem [{\citenamefont {Shaw}(1979)}]{1979_shaw_jcg}%
  \BibitemOpen
  \bibfield  {author} {\bibinfo {author} {\bibfnamefont {D.~W.}\ \bibnamefont
  {Shaw}},\ }\bibfield  {title} {\bibinfo {title} {Morphology analysis in
  localized crystal growth and dissolution},\ }\href
  {https://doi.org/https://doi.org/10.1016/0022-0248(79)90133-7} {\bibfield
  {journal} {\bibinfo  {journal} {Journal of Crystal Growth}\ }\textbf
  {\bibinfo {volume} {47}},\ \bibinfo {pages} {509} (\bibinfo {year}
  {1979})}\BibitemShut {NoStop}%
\bibitem [{\citenamefont {Herring}(1951)}]{1951_Herring_surface_theo}%
  \BibitemOpen
  \bibfield  {author} {\bibinfo {author} {\bibfnamefont {C.}~\bibnamefont
  {Herring}},\ }\bibfield  {title} {\bibinfo {title} {Some theorems on the free
  energies of crystal surfaces},\ }\href
  {https://doi.org/https://doi.org/10.1103/PhysRev.82.87} {\bibfield  {journal}
  {\bibinfo  {journal} {Physical Review}\ }\textbf {\bibinfo {volume} {82}},\
  \bibinfo {pages} {87} (\bibinfo {year} {1951})}\BibitemShut {NoStop}%
\bibitem [{\citenamefont {Cabrera}(1964)}]{1964_Cabrera_surface_theo}%
  \BibitemOpen
  \bibfield  {author} {\bibinfo {author} {\bibfnamefont {N.}~\bibnamefont
  {Cabrera}},\ }\bibfield  {title} {\bibinfo {title} {The equilibrium of
  crystal surfaces},\ }\href
  {https://doi.org/https://doi.org/10.1016/0039-6028(64)90073-1} {\bibfield
  {journal} {\bibinfo  {journal} {Surface Science}\ }\textbf {\bibinfo {volume}
  {2}},\ \bibinfo {pages} {320} (\bibinfo {year} {1964})}\BibitemShut {NoStop}%
\bibitem [{\citenamefont {Cahn}(1982)}]{1982_Cahn_gb_theo}%
  \BibitemOpen
  \bibfield  {author} {\bibinfo {author} {\bibfnamefont {J.~W.}\ \bibnamefont
  {Cahn}},\ }\bibfield  {title} {\bibinfo {title} {Transitions and phase
  equilibria among grain boundary structures},\ }\href
  {https://doi.org/https://doi.org/10.1051/jphyscol:1982619} {\bibfield
  {journal} {\bibinfo  {journal} {Journal de Physique Colloques}\ }\textbf
  {\bibinfo {volume} {22}},\ \bibinfo {pages} {199} (\bibinfo {year}
  {1982})}\BibitemShut {NoStop}%
\bibitem [{\citenamefont {Rottman}\ and\ \citenamefont
  {Wortis}(1984)}]{1984_Rottmandefacet_exp}%
  \BibitemOpen
  \bibfield  {author} {\bibinfo {author} {\bibfnamefont {C.}~\bibnamefont
  {Rottman}}\ and\ \bibinfo {author} {\bibfnamefont {M.}~\bibnamefont
  {Wortis}},\ }\bibfield  {title} {\bibinfo {title} {Equilibrium crystal shapes
  for lattice models with nearest-and next-nearest-neighbor interactions},\
  }\href {https://doi.org/https://doi.org/10.1103/PhysRevB.29.328} {\bibfield
  {journal} {\bibinfo  {journal} {Physical Review B}\ }\textbf {\bibinfo
  {volume} {29}},\ \bibinfo {pages} {328} (\bibinfo {year} {1984})}\BibitemShut
  {NoStop}%
\bibitem [{\citenamefont {Priedeman}\ and\ \citenamefont
  {Thompson}(2020)}]{2020_Priedeman_capillarity_exp}%
  \BibitemOpen
  \bibfield  {author} {\bibinfo {author} {\bibfnamefont {J.~L.}\ \bibnamefont
  {Priedeman}}\ and\ \bibinfo {author} {\bibfnamefont {G.~B.}\ \bibnamefont
  {Thompson}},\ }\bibfield  {title} {\bibinfo {title} {The influence of
  alloying in stabilizing a faceted grain boundary structure},\ }\href
  {https://doi.org/https://doi.org/10.1016/j.actamat.2020.09.085} {\bibfield
  {journal} {\bibinfo  {journal} {Acta Materialia}\ }\textbf {\bibinfo {volume}
  {201}},\ \bibinfo {pages} {329} (\bibinfo {year} {2020})}\BibitemShut
  {NoStop}%
\bibitem [{\citenamefont {Kang}\ \emph {et~al.}(2012)\citenamefont {Kang},
  \citenamefont {Wang}, \citenamefont {Zheng},\ and\ \citenamefont
  {Beyerlein}}]{2012_Kang_facetcapillarity_MD}%
  \BibitemOpen
  \bibfield  {author} {\bibinfo {author} {\bibfnamefont {K.}~\bibnamefont
  {Kang}}, \bibinfo {author} {\bibfnamefont {J.}~\bibnamefont {Wang}}, \bibinfo
  {author} {\bibfnamefont {S.~J.}\ \bibnamefont {Zheng}},\ and\ \bibinfo
  {author} {\bibfnamefont {I.~J.}\ \bibnamefont {Beyerlein}},\ }\bibfield
  {title} {\bibinfo {title} {Minimum energy structures of faceted, incoherent
  interfaces},\ }\href {https://doi.org/https://doi.org/10.1063/1.4755789}
  {\bibfield  {journal} {\bibinfo  {journal} {Journal of Applied Physics}\
  }\textbf {\bibinfo {volume} {112}},\ \bibinfo {pages} {073501} (\bibinfo
  {year} {2012})}\BibitemShut {NoStop}%
\bibitem [{\citenamefont {Kirch}\ \emph {et~al.}(2008)\citenamefont {Kirch},
  \citenamefont {Jannot}, \citenamefont {Barrales-Mora}, \citenamefont
  {Molodov},\ and\ \citenamefont
  {Gottstein}}]{2008_Kirch_facetcapillarity_halfloop_exp}%
  \BibitemOpen
  \bibfield  {author} {\bibinfo {author} {\bibfnamefont {D.~M.}\ \bibnamefont
  {Kirch}}, \bibinfo {author} {\bibfnamefont {E.}~\bibnamefont {Jannot}},
  \bibinfo {author} {\bibfnamefont {L.~A.}\ \bibnamefont {Barrales-Mora}},
  \bibinfo {author} {\bibfnamefont {D.~A.}\ \bibnamefont {Molodov}},\ and\
  \bibinfo {author} {\bibfnamefont {G.}~\bibnamefont {Gottstein}},\ }\bibfield
  {title} {\bibinfo {title} {Inclination dependence of grain boundary energy
  and its impact on the faceting and kinetics of tilt grain boundaries in
  aluminum},\ }\href
  {https://doi.org/https://doi.org/10.1016/j.actamat.2008.06.017} {\bibfield
  {journal} {\bibinfo  {journal} {Acta Materialia}\ }\textbf {\bibinfo {volume}
  {56}},\ \bibinfo {pages} {4998} (\bibinfo {year} {2008})}\BibitemShut
  {NoStop}%
\bibitem [{\citenamefont {Hanyu}\ \emph {et~al.}(2005)\citenamefont {Hanyu},
  \citenamefont {Nishimura}, \citenamefont {Matsunaga}, \citenamefont
  {Yamamoto}, \citenamefont {Ikuhara},\ and\ \citenamefont
  {Glaeser}}]{2005_Hanyu_stressfacet_exp}%
  \BibitemOpen
  \bibfield  {author} {\bibinfo {author} {\bibfnamefont {S.}~\bibnamefont
  {Hanyu}}, \bibinfo {author} {\bibfnamefont {H.}~\bibnamefont {Nishimura}},
  \bibinfo {author} {\bibfnamefont {K.}~\bibnamefont {Matsunaga}}, \bibinfo
  {author} {\bibfnamefont {T.}~\bibnamefont {Yamamoto}}, \bibinfo {author}
  {\bibfnamefont {Y.}~\bibnamefont {Ikuhara}},\ and\ \bibinfo {author}
  {\bibfnamefont {A.~M.}\ \bibnamefont {Glaeser}},\ }\bibfield  {title}
  {\bibinfo {title} {Stress-induced facet coarsening in a {$\Sigma$} 7
  {[4$\bar{5}$10]} symmetrical tilt grain boundary in an alumina bicrystal},\
  }\href {https://doi.org/https://doi.org/10.1007/s10853-005-2675-3} {\bibfield
   {journal} {\bibinfo  {journal} {Journal of Materials Science}\ }\textbf
  {\bibinfo {volume} {40}},\ \bibinfo {pages} {3137–3142} (\bibinfo {year}
  {2005})}\BibitemShut {NoStop}%
\bibitem [{\citenamefont {Hirth}\ \emph {et~al.}(2020)\citenamefont {Hirth},
  \citenamefont {Hirth},\ and\ \citenamefont
  {Wang}}]{2020_Hirth_disconnection_review}%
  \BibitemOpen
  \bibfield  {author} {\bibinfo {author} {\bibfnamefont {J.~P.}\ \bibnamefont
  {Hirth}}, \bibinfo {author} {\bibfnamefont {G.}~\bibnamefont {Hirth}},\ and\
  \bibinfo {author} {\bibfnamefont {J.}~\bibnamefont {Wang}},\ }\bibfield
  {title} {\bibinfo {title} {Disclinations and disconnections in minerals and
  metals},\ }\href {https://doi.org/https://doi.org/10.1073/pnas.1915140117}
  {\bibfield  {journal} {\bibinfo  {journal} {Proceedings of the National
  Academy of Sciences}\ }\textbf {\bibinfo {volume} {117}},\ \bibinfo {pages}
  {196} (\bibinfo {year} {2020})}\BibitemShut {NoStop}%
\bibitem [{\citenamefont {Han}\ \emph {et~al.}(2018)\citenamefont {Han},
  \citenamefont {Thomas},\ and\ \citenamefont
  {Srolovitz}}]{2018_Han_gbkinetics_review}%
  \BibitemOpen
  \bibfield  {author} {\bibinfo {author} {\bibfnamefont {J.}~\bibnamefont
  {Han}}, \bibinfo {author} {\bibfnamefont {S.~L.}\ \bibnamefont {Thomas}},\
  and\ \bibinfo {author} {\bibfnamefont {D.~J.}\ \bibnamefont {Srolovitz}},\
  }\bibfield  {title} {\bibinfo {title} {Grain-boundary kinetics: {A} unified
  approach},\ }\href
  {https://doi.org/https://doi.org/10.1016/j.pmatsci.2018.05.004} {\bibfield
  {journal} {\bibinfo  {journal} {Progress in Materials Science}\ }\textbf
  {\bibinfo {volume} {98}},\ \bibinfo {pages} {386} (\bibinfo {year}
  {2018})}\BibitemShut {NoStop}%
\bibitem [{\citenamefont {Rajabzadeh}\ \emph {et~al.}(2013)\citenamefont
  {Rajabzadeh}, \citenamefont {Legros}, \citenamefont {Combe}, \citenamefont
  {Mompiou},\ and\ \citenamefont
  {Molodov}}]{2013_Rajabzadeh_disconnection_exp}%
  \BibitemOpen
  \bibfield  {author} {\bibinfo {author} {\bibfnamefont {A.}~\bibnamefont
  {Rajabzadeh}}, \bibinfo {author} {\bibfnamefont {M.}~\bibnamefont {Legros}},
  \bibinfo {author} {\bibfnamefont {N.}~\bibnamefont {Combe}}, \bibinfo
  {author} {\bibfnamefont {F.}~\bibnamefont {Mompiou}},\ and\ \bibinfo {author}
  {\bibfnamefont {D.~A.}\ \bibnamefont {Molodov}},\ }\bibfield  {title}
  {\bibinfo {title} {Evidence of grain boundary dislocation step motion
  associated to shear-coupled grain boundary migration},\ }\href
  {https://doi.org/https://doi.org/10.1080/14786435.2012.760760} {\bibfield
  {journal} {\bibinfo  {journal} {Philosophical Magazine}\ }\textbf {\bibinfo
  {volume} {93}},\ \bibinfo {pages} {1299} (\bibinfo {year}
  {2013})}\BibitemShut {NoStop}%
\bibitem [{\citenamefont {Hamilton}\ \emph {et~al.}(2003)\citenamefont
  {Hamilton}, \citenamefont {Siegel}, \citenamefont {Daruka},\ and\
  \citenamefont {Léonard}}]{2003_Hamilton_stresseffect_theo}%
  \BibitemOpen
  \bibfield  {author} {\bibinfo {author} {\bibfnamefont {J.~C.}\ \bibnamefont
  {Hamilton}}, \bibinfo {author} {\bibfnamefont {D.~J.}\ \bibnamefont
  {Siegel}}, \bibinfo {author} {\bibfnamefont {I.}~\bibnamefont {Daruka}},\
  and\ \bibinfo {author} {\bibfnamefont {F.}~\bibnamefont {Léonard}},\
  }\bibfield  {title} {\bibinfo {title} {Why do grain boundaries exhibit finite
  facet lengths?},\ }\href
  {https://doi.org/https://doi.org/10.1103/PhysRevLett.90.246102} {\bibfield
  {journal} {\bibinfo  {journal} {Physical Review Letters}\ }\textbf {\bibinfo
  {volume} {90}},\ \bibinfo {pages} {246102} (\bibinfo {year}
  {2003})}\BibitemShut {NoStop}%
\bibitem [{\citenamefont {Wu}\ \emph {et~al.}(2009)\citenamefont {Wu},
  \citenamefont {Zhang},\ and\ \citenamefont
  {Srolovitz}}]{2009_Wu_stresseffect_MD}%
  \BibitemOpen
  \bibfield  {author} {\bibinfo {author} {\bibfnamefont {Z.~X.}\ \bibnamefont
  {Wu}}, \bibinfo {author} {\bibfnamefont {Y.~W.}\ \bibnamefont {Zhang}},\ and\
  \bibinfo {author} {\bibfnamefont {D.~J.}\ \bibnamefont {Srolovitz}},\
  }\bibfield  {title} {\bibinfo {title} {Grain boundary finite length
  faceting},\ }\href
  {https://doi.org/https://doi.org/10.1016/j.actamat.2009.05.026} {\bibfield
  {journal} {\bibinfo  {journal} {Acta Materialia}\ }\textbf {\bibinfo {volume}
  {57}},\ \bibinfo {pages} {4278} (\bibinfo {year} {2009})}\BibitemShut
  {NoStop}%
\bibitem [{\citenamefont {Medlin}\ \emph {et~al.}(2017)\citenamefont {Medlin},
  \citenamefont {Hattar}, \citenamefont {Zimmerman}, \citenamefont
  {Abdeljawad},\ and\ \citenamefont {Foiles}}]{2017_Medlin_stresseffect_MD}%
  \BibitemOpen
  \bibfield  {author} {\bibinfo {author} {\bibfnamefont {D.~L.}\ \bibnamefont
  {Medlin}}, \bibinfo {author} {\bibfnamefont {K.}~\bibnamefont {Hattar}},
  \bibinfo {author} {\bibfnamefont {J.~A.}\ \bibnamefont {Zimmerman}}, \bibinfo
  {author} {\bibfnamefont {F.}~\bibnamefont {Abdeljawad}},\ and\ \bibinfo
  {author} {\bibfnamefont {S.~M.}\ \bibnamefont {Foiles}},\ }\bibfield  {title}
  {\bibinfo {title} {Defect character at grain boundary facet junctions:
  Analysis of an asymmetric {$\Sigma$} = 5 grain boundary in {F}e},\ }\href
  {https://doi.org/https://doi.org/10.1016/j.actamat.2009.05.026} {\bibfield
  {journal} {\bibinfo  {journal} {Acta Materialia}\ }\textbf {\bibinfo {volume}
  {124}},\ \bibinfo {pages} {383} (\bibinfo {year} {2017})}\BibitemShut
  {NoStop}%
\bibitem [{\citenamefont {Han}\ \emph {et~al.}(2022)\citenamefont {Han},
  \citenamefont {Srolovitz},\ and\ \citenamefont
  {Salvalaglio}}]{2022_Han_EOM_theory}%
  \BibitemOpen
  \bibfield  {author} {\bibinfo {author} {\bibfnamefont {J.}~\bibnamefont
  {Han}}, \bibinfo {author} {\bibfnamefont {D.~J.}\ \bibnamefont {Srolovitz}},\
  and\ \bibinfo {author} {\bibfnamefont {M.}~\bibnamefont {Salvalaglio}},\
  }\bibfield  {title} {\bibinfo {title} {Disconnection-mediated migration of
  interfaces in microstructures: I. continuum model},\ }\href
  {https://doi.org/https://doi.org/10.1016/j.actamat.2021.117178} {\bibfield
  {journal} {\bibinfo  {journal} {Acta Materialia}\ }\textbf {\bibinfo {volume}
  {227}},\ \bibinfo {pages} {117178} (\bibinfo {year} {2022})}\BibitemShut
  {NoStop}%
\bibitem [{\citenamefont {Salvalaglio}\ \emph {et~al.}(2022)\citenamefont
  {Salvalaglio}, \citenamefont {Srolovitz},\ and\ \citenamefont
  {Han}}]{2022_Marco_diffuse_PF}%
  \BibitemOpen
  \bibfield  {author} {\bibinfo {author} {\bibfnamefont {M.}~\bibnamefont
  {Salvalaglio}}, \bibinfo {author} {\bibfnamefont {D.~J.}\ \bibnamefont
  {Srolovitz}},\ and\ \bibinfo {author} {\bibfnamefont {J.}~\bibnamefont
  {Han}},\ }\bibfield  {title} {\bibinfo {title} {Disconnection-mediated
  migration of interfaces in microstructures: {II}. diffuse interface
  simulations},\ }\href
  {https://doi.org/https://doi.org/10.1016/j.actamat.2021.117463} {\bibfield
  {journal} {\bibinfo  {journal} {Acta Materialia}\ }\textbf {\bibinfo {volume}
  {227}},\ \bibinfo {pages} {117463} (\bibinfo {year} {2022})}\BibitemShut
  {NoStop}%
\bibitem [{\citenamefont {Ranganathan}(1966)}]{1966_Ranganathan_CSL_theory}%
  \BibitemOpen
  \bibfield  {author} {\bibinfo {author} {\bibfnamefont {S.}~\bibnamefont
  {Ranganathan}},\ }\bibfield  {title} {\bibinfo {title} {On the geometry of
  coincidence-site lattices},\ }\href
  {https://doi.org/https://doi.org/10.1107/S0365110X66002615} {\bibfield
  {journal} {\bibinfo  {journal} {Acta Crystallographica}\ }\textbf {\bibinfo
  {volume} {21}},\ \bibinfo {pages} {197} (\bibinfo {year} {1966})}\BibitemShut
  {NoStop}%
\bibitem [{\citenamefont {Bollmann}(1966)}]{1966_Bollmann_CSL_theory}%
  \BibitemOpen
  \bibfield  {author} {\bibinfo {author} {\bibfnamefont {W.}~\bibnamefont
  {Bollmann}},\ }\bibfield  {title} {\bibinfo {title} {On the geometry of grain
  and phase boundaries: I. general theory},\ }\href
  {https://doi.org/https://doi.org/10.1080/14786436708229748} {\bibfield
  {journal} {\bibinfo  {journal} {The Philosophical Magazine: A Journal of
  Theoretical Experimental and Applied Physics}\ }\textbf {\bibinfo {volume}
  {16}},\ \bibinfo {pages} {363} (\bibinfo {year} {1966})}\BibitemShut
  {NoStop}%
\bibitem [{\citenamefont {Hirth}\ and\ \citenamefont
  {Balluffi}(1973)}]{1997_Hirth_CSL_theory}%
  \BibitemOpen
  \bibfield  {author} {\bibinfo {author} {\bibfnamefont {J.~P.}\ \bibnamefont
  {Hirth}}\ and\ \bibinfo {author} {\bibfnamefont {R.~W.}\ \bibnamefont
  {Balluffi}},\ }\bibfield  {title} {\bibinfo {title} {On grain boundary
  dislocations and ledges},\ }\href
  {https://doi.org/https://doi.org/10.1016/0001-6160(73)90150-8} {\bibfield
  {journal} {\bibinfo  {journal} {Acta Metallurgica}\ }\textbf {\bibinfo
  {volume} {21}},\ \bibinfo {pages} {929} (\bibinfo {year} {1973})}\BibitemShut
  {NoStop}%
\bibitem [{\citenamefont {Zhang}\ \emph {et~al.}(2017)\citenamefont {Zhang},
  \citenamefont {Han}, \citenamefont {Xiang},\ and\ \citenamefont
  {Srolovitz}}]{2017_Zhang_EOM_theory}%
  \BibitemOpen
  \bibfield  {author} {\bibinfo {author} {\bibfnamefont {L.}~\bibnamefont
  {Zhang}}, \bibinfo {author} {\bibfnamefont {J.}~\bibnamefont {Han}}, \bibinfo
  {author} {\bibfnamefont {Y.}~\bibnamefont {Xiang}},\ and\ \bibinfo {author}
  {\bibfnamefont {D.~J.}\ \bibnamefont {Srolovitz}},\ }\bibfield  {title}
  {\bibinfo {title} {Equation of motion for a grain boundary},\ }\href
  {https://doi.org/https://doi.org/10.1080/14786436708229748} {\bibfield
  {journal} {\bibinfo  {journal} {Physical Review Letters}\ }\textbf {\bibinfo
  {volume} {119}},\ \bibinfo {pages} {246101} (\bibinfo {year}
  {2017})}\BibitemShut {NoStop}%
\bibitem [{\citenamefont {Radetic}\ \emph {et~al.}(2011)\citenamefont
  {Radetic}, \citenamefont {Minor},\ and\ \citenamefont
  {Dahmen}}]{2011_Radetic_squareAu_exp}%
  \BibitemOpen
  \bibfield  {author} {\bibinfo {author} {\bibfnamefont {T.}~\bibnamefont
  {Radetic}}, \bibinfo {author} {\bibfnamefont {A.~M.}\ \bibnamefont {Minor}},\
  and\ \bibinfo {author} {\bibfnamefont {U.}~\bibnamefont {Dahmen}},\
  }\bibfield  {title} {\bibinfo {title} {Capillarity-driven migration of a thin
  {G}e wedge in contact with a bicrystalline {A}u film},\ }\href
  {https://doi.org/https://doi.org/10.1016/j.actamat.2010.12.051} {\bibfield
  {journal} {\bibinfo  {journal} {Acta Materialia}\ }\textbf {\bibinfo {volume}
  {59}},\ \bibinfo {pages} {2481} (\bibinfo {year} {2011})}\BibitemShut
  {NoStop}%
\bibitem [{\citenamefont {Würschum}\ and\ \citenamefont
  {Balluffi}(1993)}]{1996_Balluffi_squaredisconnection_exp}%
  \BibitemOpen
  \bibfield  {author} {\bibinfo {author} {\bibfnamefont {R.}~\bibnamefont
  {Würschum}}\ and\ \bibinfo {author} {\bibfnamefont {R.~W.}\ \bibnamefont
  {Balluffi}},\ }\bibfield  {title} {\bibinfo {title} {In‐situ study of the
  migration of grain boundary facets in {A}u bicrystals under high driving
  forces},\ }\href {https://doi.org/https://doi.org/10.1002/pssa.2211360206}
  {\bibfield  {journal} {\bibinfo  {journal} {physica status solidi (a)}\
  }\textbf {\bibinfo {volume} {136}},\ \bibinfo {pages} {323} (\bibinfo {year}
  {1993})}\BibitemShut {NoStop}%
\bibitem [{\citenamefont {Radetic}\ \emph {et~al.}(2012)\citenamefont
  {Radetic}, \citenamefont {Ophus}, \citenamefont {Olmsted}, \citenamefont
  {Asta},\ and\ \citenamefont {Dahmen}}]{2012_Radetic_squareAu_exp}%
  \BibitemOpen
  \bibfield  {author} {\bibinfo {author} {\bibfnamefont {T.}~\bibnamefont
  {Radetic}}, \bibinfo {author} {\bibfnamefont {C.}~\bibnamefont {Ophus}},
  \bibinfo {author} {\bibfnamefont {D.~L.}\ \bibnamefont {Olmsted}}, \bibinfo
  {author} {\bibfnamefont {M.}~\bibnamefont {Asta}},\ and\ \bibinfo {author}
  {\bibfnamefont {U.}~\bibnamefont {Dahmen}},\ }\bibfield  {title} {\bibinfo
  {title} {Mechanism and dynamics of shrinking island grains in mazed bicrystal
  thin films of {A}u},\ }\href
  {https://doi.org/https://doi.org/10.1016/j.actamat.2012.09.012} {\bibfield
  {journal} {\bibinfo  {journal} {Acta Materialia}\ }\textbf {\bibinfo {volume}
  {60}},\ \bibinfo {pages} {7051} (\bibinfo {year} {2012})}\BibitemShut
  {NoStop}%
\bibitem [{\citenamefont {Gruber}\ and\ \citenamefont
  {Mullins}(1967)}]{1967_Gruber_TLK_theory}%
  \BibitemOpen
  \bibfield  {author} {\bibinfo {author} {\bibfnamefont {E.~E.}\ \bibnamefont
  {Gruber}}\ and\ \bibinfo {author} {\bibfnamefont {W.~W.}\ \bibnamefont
  {Mullins}},\ }\bibfield  {title} {\bibinfo {title} {On the theory of
  anisotropy of crystalline surface tension},\ }\href
  {https://doi.org/https://doi.org/10.1016/0022-3697(67)90017-0} {\bibfield
  {journal} {\bibinfo  {journal} {Journal of Physics and Chemistry of Solids}\
  }\textbf {\bibinfo {volume} {28}},\ \bibinfo {pages} {875} (\bibinfo {year}
  {1967})}\BibitemShut {NoStop}%
\bibitem [{\citenamefont {Tschopp}\ and\ \citenamefont
  {Mcdowell}(2007)}]{2007_McDowell_GBenergy_MD}%
  \BibitemOpen
  \bibfield  {author} {\bibinfo {author} {\bibfnamefont {M.~A.}\ \bibnamefont
  {Tschopp}}\ and\ \bibinfo {author} {\bibfnamefont {D.~L.}\ \bibnamefont
  {Mcdowell}},\ }\bibfield  {title} {\bibinfo {title} {Asymmetric tilt grain
  boundary structure and energy in {C}opper and {A}luminium},\ }\href
  {https://doi.org/https://doi.org/10.1080/14786430701455321} {\bibfield
  {journal} {\bibinfo  {journal} {Philosophical Magazine}\ }\textbf {\bibinfo
  {volume} {87}},\ \bibinfo {pages} {3871} (\bibinfo {year}
  {2007})}\BibitemShut {NoStop}%
\bibitem [{\citenamefont {Burton}\ \emph {et~al.}(1951)\citenamefont {Burton},
  \citenamefont {Cabrera},\ and\ \citenamefont
  {Frank}}]{1951_Burton_surfaceenergy_theory}%
  \BibitemOpen
  \bibfield  {author} {\bibinfo {author} {\bibfnamefont {W.~K.}\ \bibnamefont
  {Burton}}, \bibinfo {author} {\bibfnamefont {N.}~\bibnamefont {Cabrera}},\
  and\ \bibinfo {author} {\bibfnamefont {F.~C.}\ \bibnamefont {Frank}},\
  }\bibfield  {title} {\bibinfo {title} {The growth of crystals and the
  equilibrium structure of their surfaces},\ }\href
  {https://doi.org/https://doi.org/10.1098/rsta.1951.0006} {\bibfield
  {journal} {\bibinfo  {journal} {Philosophical Transactions of the Royal
  Society of London. Series A, Mathematical and Physical Sciences}\ }\textbf
  {\bibinfo {volume} {243}},\ \bibinfo {pages} {299–358} (\bibinfo {year}
  {1951})}\BibitemShut {NoStop}%
\bibitem [{\citenamefont {Avron}\ \emph {et~al.}(1982)\citenamefont {Avron},
  \citenamefont {van Beijeren}, \citenamefont {Schulman},\ and\ \citenamefont
  {Zia}}]{1982_Avron_SOS_theory}%
  \BibitemOpen
  \bibfield  {author} {\bibinfo {author} {\bibfnamefont {J.~E.}\ \bibnamefont
  {Avron}}, \bibinfo {author} {\bibfnamefont {H.}~\bibnamefont {van Beijeren}},
  \bibinfo {author} {\bibfnamefont {L.~S.}\ \bibnamefont {Schulman}},\ and\
  \bibinfo {author} {\bibfnamefont {R.~K.~P.}\ \bibnamefont {Zia}},\ }\bibfield
   {title} {\bibinfo {title} {Roughening transition, surface tension and
  equilibrium droplet shapes in a two-dimensional {I}sing system},\ }\href
  {https://doi.org/https://doi.org/10.1088/0305-4470/15/2/006} {\bibfield
  {journal} {\bibinfo  {journal} {Journal of Physics A: Mathematical and
  General}\ }\textbf {\bibinfo {volume} {15}},\ \bibinfo {pages} {L81}
  (\bibinfo {year} {1982})}\BibitemShut {NoStop}%
\bibitem [{\citenamefont {Barrales-Mor}\ and\ \citenamefont
  {Molodov}(2016)}]{2016_Molodov_circular_MD}%
  \BibitemOpen
  \bibfield  {author} {\bibinfo {author} {\bibfnamefont {L.~A.}\ \bibnamefont
  {Barrales-Mor}}\ and\ \bibinfo {author} {\bibfnamefont {D.~A.}\ \bibnamefont
  {Molodov}},\ }\bibfield  {title} {\bibinfo {title} {Capillarity-driven
  shrinkage of grains with tilt and mixed boundaries studied by molecular
  dynamics},\ }\href
  {https://doi.org/https://doi.org/10.1016/j.actamat.2016.08.060} {\bibfield
  {journal} {\bibinfo  {journal} {Acta Materialia}\ }\textbf {\bibinfo {volume}
  {120}},\ \bibinfo {pages} {179} (\bibinfo {year} {2016})}\BibitemShut
  {NoStop}%
\bibitem [{\citenamefont {Brandenburg}\ \emph {et~al.}(2014)\citenamefont
  {Brandenburg}, \citenamefont {Barrales-Mora},\ and\ \citenamefont
  {Molodov}}]{2014_Molodov_circular_MD}%
  \BibitemOpen
  \bibfield  {author} {\bibinfo {author} {\bibfnamefont {J.~E.}\ \bibnamefont
  {Brandenburg}}, \bibinfo {author} {\bibfnamefont {L.~A.}\ \bibnamefont
  {Barrales-Mora}},\ and\ \bibinfo {author} {\bibfnamefont {D.~A.}\
  \bibnamefont {Molodov}},\ }\bibfield  {title} {\bibinfo {title} {On migration
  and faceting of low-angle grain boundaries: Experimental and computational
  study},\ }\href
  {https://doi.org/https://doi.org/10.1016/j.actamat.2014.06.006} {\bibfield
  {journal} {\bibinfo  {journal} {Acta Materialia}\ }\textbf {\bibinfo {volume}
  {77}},\ \bibinfo {pages} {294} (\bibinfo {year} {2014})}\BibitemShut
  {NoStop}%
\bibitem [{\citenamefont {Du}\ \emph {et~al.}(2005)\citenamefont {Du},
  \citenamefont {Srolovitz}, \citenamefont {Coltrin},\ and\ \citenamefont
  {Mitchell}}]{2005_Du_vplot_theo}%
  \BibitemOpen
  \bibfield  {author} {\bibinfo {author} {\bibfnamefont {D.}~\bibnamefont
  {Du}}, \bibinfo {author} {\bibfnamefont {D.~J.}\ \bibnamefont {Srolovitz}},
  \bibinfo {author} {\bibfnamefont {M.~E.}\ \bibnamefont {Coltrin}},\ and\
  \bibinfo {author} {\bibfnamefont {C.~C.}\ \bibnamefont {Mitchell}},\
  }\bibfield  {title} {\bibinfo {title} {Systematic prediction of kinetically
  limited crystal growth morphologies},\ }\href
  {https://doi.org/https://doi.org/10.1103/PhysRevLett.95.155503} {\bibfield
  {journal} {\bibinfo  {journal} {Physical Review Letters}\ }\textbf {\bibinfo
  {volume} {95}},\ \bibinfo {pages} {155503} (\bibinfo {year}
  {2005})}\BibitemShut {NoStop}%
\bibitem [{\citenamefont {Mompiou}\ \emph {et~al.}(2009)\citenamefont
  {Mompiou}, \citenamefont {Caillard},\ and\ \citenamefont
  {Legros}}]{2009_Monpiou_shearcoupling_exp}%
  \BibitemOpen
  \bibfield  {author} {\bibinfo {author} {\bibfnamefont {F.}~\bibnamefont
  {Mompiou}}, \bibinfo {author} {\bibfnamefont {D.}~\bibnamefont {Caillard}},\
  and\ \bibinfo {author} {\bibfnamefont {M.}~\bibnamefont {Legros}},\
  }\bibfield  {title} {\bibinfo {title} {Grain boundary shear–migration
  coupling—{I}. {I}n-situ {TEM} straining experiments in al polycrystals},\
  }\href {https://doi.org/https://doi.org/10.1016/j.actamat.2009.01.014}
  {\bibfield  {journal} {\bibinfo  {journal} {Acta Materialia}\ }\textbf
  {\bibinfo {volume} {57}},\ \bibinfo {pages} {2198} (\bibinfo {year}
  {2009})}\BibitemShut {NoStop}%
\bibitem [{\citenamefont {Trautt}\ \emph {et~al.}(2012)\citenamefont {Trautt},
  \citenamefont {Adland}, \citenamefont {Karma},\ and\ \citenamefont
  {Mishin}}]{2012_Trautt_shearcoupling_MDPFC}%
  \BibitemOpen
  \bibfield  {author} {\bibinfo {author} {\bibfnamefont {Z.~T.}\ \bibnamefont
  {Trautt}}, \bibinfo {author} {\bibfnamefont {A.}~\bibnamefont {Adland}},
  \bibinfo {author} {\bibfnamefont {A.}~\bibnamefont {Karma}},\ and\ \bibinfo
  {author} {\bibfnamefont {Y.}~\bibnamefont {Mishin}},\ }\bibfield  {title}
  {\bibinfo {title} {Coupled motion of asymmetrical tilt grain boundaries:
  Molecular dynamics and phase field crystal simulations},\ }\href
  {https://doi.org/https://doi.org/10.1016/j.actamat.2012.08.018} {\bibfield
  {journal} {\bibinfo  {journal} {Acta Materialia}\ }\textbf {\bibinfo {volume}
  {60}},\ \bibinfo {pages} {6528} (\bibinfo {year} {2012})}\BibitemShut
  {NoStop}%
\bibitem [{\citenamefont {Gottstein}\ and\ \citenamefont
  {Shvindlerman}(1999)}]{halfloop_book}%
  \BibitemOpen
  \bibfield  {author} {\bibinfo {author} {\bibfnamefont {G.}~\bibnamefont
  {Gottstein}}\ and\ \bibinfo {author} {\bibfnamefont {L.~S.}\ \bibnamefont
  {Shvindlerman}},\ }\href@noop {} {\emph {\bibinfo {title} {Grain boundary
  migration in metals: thermodynamics, kinetics, applications}}}\ (\bibinfo
  {publisher} {CRC Press},\ \bibinfo {year} {1999})\BibitemShut {NoStop}%
\bibitem [{\citenamefont {Aristov}\ \emph {et~al.}(1973)\citenamefont
  {Aristov}, \citenamefont {Fridman},\ and\ \citenamefont
  {Shvindlerman}}]{1973_Aristov_halfloop_exp}%
  \BibitemOpen
  \bibfield  {author} {\bibinfo {author} {\bibfnamefont {V.~Y.}\ \bibnamefont
  {Aristov}}, \bibinfo {author} {\bibfnamefont {E.~M.}\ \bibnamefont
  {Fridman}},\ and\ \bibinfo {author} {\bibfnamefont {L.~S.}\ \bibnamefont
  {Shvindlerman}},\ }\bibfield  {title} {\bibinfo {title} {Boundary migration
  in aluminum bicrystals},\ }\href@noop {} {\bibfield  {journal} {\bibinfo
  {journal} {Phys. Met. Metallogr.}\ }\textbf {\bibinfo {volume} {35}},\
  \bibinfo {pages} {187} (\bibinfo {year} {1973})}\BibitemShut {NoStop}%
\bibitem [{\citenamefont {Straumal}\ \emph {et~al.}(2008)\citenamefont
  {Straumal}, \citenamefont {Gornakova},\ and\ \citenamefont
  {Sursaeva}}]{2008_Straumal_Zinc1_exp}%
  \BibitemOpen
  \bibfield  {author} {\bibinfo {author} {\bibfnamefont {B.~B.}\ \bibnamefont
  {Straumal}}, \bibinfo {author} {\bibfnamefont {A.~S.}\ \bibnamefont
  {Gornakova}},\ and\ \bibinfo {author} {\bibfnamefont {V.~G.}\ \bibnamefont
  {Sursaeva}},\ }\bibfield  {title} {\bibinfo {title} {Reversible
  transformation of a grain-boundary facet into a rough-to-rough ridge in
  {Z}inc},\ }\href {https://doi.org/https://doi.org/10.1080/09500830701716967}
  {\bibfield  {journal} {\bibinfo  {journal} {Philosophical Magazine Letters}\
  }\textbf {\bibinfo {volume} {88}},\ \bibinfo {pages} {27} (\bibinfo {year}
  {2008})}\BibitemShut {NoStop}%
\bibitem [{\citenamefont {Sursaeva}\ \emph {et~al.}(2008)\citenamefont
  {Sursaeva}, \citenamefont {B.Straumal}, \citenamefont {Gornakova},
  \citenamefont {Shvindlerman},\ and\ \citenamefont
  {Gottstein}}]{2008_Sursaeva_Zinc2_exp}%
  \BibitemOpen
  \bibfield  {author} {\bibinfo {author} {\bibfnamefont {V.~G.}\ \bibnamefont
  {Sursaeva}}, \bibinfo {author} {\bibfnamefont {B.}~\bibnamefont
  {B.Straumal}}, \bibinfo {author} {\bibfnamefont {A.~S.}\ \bibnamefont
  {Gornakova}}, \bibinfo {author} {\bibfnamefont {L.~S.}\ \bibnamefont
  {Shvindlerman}},\ and\ \bibinfo {author} {\bibfnamefont {G.}~\bibnamefont
  {Gottstein}},\ }\bibfield  {title} {\bibinfo {title} {Effect of faceting on
  grain boundary motion in {Z}n},\ }\href
  {https://doi.org/https://doi.org/10.1016/j.actamat.2008.02.014} {\bibfield
  {journal} {\bibinfo  {journal} {Acta Materialia}\ }\textbf {\bibinfo {volume}
  {56}},\ \bibinfo {pages} {2728} (\bibinfo {year} {2008})}\BibitemShut
  {NoStop}%
\bibitem [{\citenamefont {Upmanyu}\ \emph {et~al.}(1998)\citenamefont
  {Upmanyu}, \citenamefont {Smith},\ and\ \citenamefont
  {Srolovitz}}]{1998_Upmanyu_halfloopAl_MD}%
  \BibitemOpen
  \bibfield  {author} {\bibinfo {author} {\bibfnamefont {M.}~\bibnamefont
  {Upmanyu}}, \bibinfo {author} {\bibfnamefont {R.}~\bibnamefont {Smith}},\
  and\ \bibinfo {author} {\bibfnamefont {D.~J.}\ \bibnamefont {Srolovitz}},\
  }\bibfield  {title} {\bibinfo {title} {Atomistic simulation of curvature
  driven grain boundary migration},\ }\href
  {https://doi.org/https://doi.org/10.1023/A:1008608418845} {\bibfield
  {journal} {\bibinfo  {journal} {Interface Science}\ }\textbf {\bibinfo
  {volume} {6}},\ \bibinfo {pages} {41–58} (\bibinfo {year}
  {1998})}\BibitemShut {NoStop}%
\bibitem [{\citenamefont {Hsieh}\ and\ \citenamefont
  {Balluffi}(1989)}]{1989_Balluffi_defaceting_exp}%
  \BibitemOpen
  \bibfield  {author} {\bibinfo {author} {\bibfnamefont {T.~E.}\ \bibnamefont
  {Hsieh}}\ and\ \bibinfo {author} {\bibfnamefont {R.~W.}\ \bibnamefont
  {Balluffi}},\ }\bibfield  {title} {\bibinfo {title} {Observations of
  roughening/de-faceting phase transitions in grain boundaries},\ }\href
  {https://doi.org/https://doi.org/10.1016/0001-6160(89)90138-7} {\bibfield
  {journal} {\bibinfo  {journal} {Acta Metallurgica}\ }\textbf {\bibinfo
  {volume} {37}},\ \bibinfo {pages} {2133} (\bibinfo {year}
  {1989})}\BibitemShut {NoStop}%
\bibitem [{\citenamefont {Hausser}\ and\ \citenamefont
  {Voigt}(2005)}]{hausser2005facet}%
  \BibitemOpen
  \bibfield  {author} {\bibinfo {author} {\bibfnamefont {F.}~\bibnamefont
  {Hausser}}\ and\ \bibinfo {author} {\bibfnamefont {A.}~\bibnamefont
  {Voigt}},\ }\bibfield  {title} {\bibinfo {title} {Facet formation and
  coarsening modeled by a geometric evolution law for epitaxial growth},\
  }\href {https://doi.org/https://doi.org/10.1016/j.jcrysgro.2004.10.137}
  {\bibfield  {journal} {\bibinfo  {journal} {Journal of Crystal Growth}\
  }\textbf {\bibinfo {volume} {275}},\ \bibinfo {pages} {e47} (\bibinfo {year}
  {2005})}\BibitemShut {NoStop}%
\bibitem [{\citenamefont {Hamilton}(2006)}]{hamilton2006}%
  \BibitemOpen
  \bibfield  {author} {\bibinfo {author} {\bibfnamefont {J.~C.}\ \bibnamefont
  {Hamilton}},\ }\bibfield  {title} {\bibinfo {title} {Edge energies: Atomistic
  calculations of a continuum quantity},\ }\href
  {https://doi.org/10.1103/PhysRevB.73.125447} {\bibfield  {journal} {\bibinfo
  {journal} {Phys. Rev. B}\ }\textbf {\bibinfo {volume} {73}},\ \bibinfo
  {pages} {125447} (\bibinfo {year} {2006})}\BibitemShut {NoStop}%
\bibitem [{\citenamefont {Chen}\ \emph {et~al.}(2019)\citenamefont {Chen},
  \citenamefont {Han}, \citenamefont {Thomas},\ and\ \citenamefont
  {Srolovitz}}]{2019_Chen_disconnection_MC}%
  \BibitemOpen
  \bibfield  {author} {\bibinfo {author} {\bibfnamefont {K.}~\bibnamefont
  {Chen}}, \bibinfo {author} {\bibfnamefont {J.}~\bibnamefont {Han}}, \bibinfo
  {author} {\bibfnamefont {S.~L.}\ \bibnamefont {Thomas}},\ and\ \bibinfo
  {author} {\bibfnamefont {D.~J.}\ \bibnamefont {Srolovitz}},\ }\bibfield
  {title} {\bibinfo {title} {Grain boundary shear coupling is not a grain
  boundary property},\ }\href
  {https://doi.org/https://doi.org/10.1016/j.actamat.2019.01.040} {\bibfield
  {journal} {\bibinfo  {journal} {Acta Materialia}\ }\textbf {\bibinfo {volume}
  {167}},\ \bibinfo {pages} {241} (\bibinfo {year} {2019})}\BibitemShut
  {NoStop}%
\end{thebibliography}%

\end{document}